\documentclass[aps,prx,showpacs,twocolumn]{revtex4-2}
\usepackage{graphicx}
\usepackage{subfig}
\usepackage{tabularx}
\usepackage{bm}
\usepackage{amsmath}
\usepackage{esvect}

\providecommand{\bra}[1]{\langle #1 \rvert}
\providecommand{\ket}[1]{\lvert #1 \rangle}

\providecommand{\ketbra}[2]{\lvert  #1\rangle \langle #2 \rvert}
\providecommand{\be}{\begin{equation}}
\providecommand{\ee}{\end{equation}}
\providecommand{\ba}{\begin{eqnarray}}
\providecommand{\ea}{\end{eqnarray}}

\newcommand{\Lim}[1]{\raisebox{0.5ex}{\scalebox{0.8}{$\displaystyle \lim_{#1}\;$}}}

\usepackage{mathbbol}

\usepackage{tikz}
\usepackage{amsfonts,amssymb}
\usepackage{dsfont}
\usepackage{physics}
\usepackage{tocvsec2}

\usepackage{graphicx}
\usepackage{subfig}
\usepackage{tabularx}
\usepackage{bm}
\usepackage{amsmath}
\usepackage{esvect}

\usepackage{siunitx}
\usepackage{hyperref}
\usepackage[capitalize]{cleveref}

\providecommand{\bra}[1]{\langle #1 \rvert}
\providecommand{\ket}[1]{\lvert #1 \rangle}

\providecommand{\ketbra}[2]{\lvert  #1\rangle \langle #2 \rvert}
\providecommand{\be}{\begin{equation}}
\providecommand{\ee}{\end{equation}}
\providecommand{\ba}{\begin{eqnarray}}
\providecommand{\ea}{\end{eqnarray}}
\usepackage{amsfonts,amssymb}
\usepackage{dsfont}
\usepackage{physics}
\usepackage{tocvsec2}

\hypersetup{colorlinks=true,linkcolor=blue,citecolor = blue, urlcolor=black}
\newcommand{\beq}{\begin{equation}}
\newcommand{\eeq}{\end{equation}}

\usepackage{appendix}

\begin{document}

\title{Unified framework for bosonic quantum information encoding, resources and universality from superselection rules}

\author{Eloi Descamps$^{1}$}
\author{Astghik Saharyan$^{1}$}
\author{Adrien Chivet$^{1,2}$}
\author{ Arne Keller$^{1,3}$}
\author{Pérola Milman$^{1}$ }
\email{corresponding author: perola.milman@u-paris.fr}

\affiliation{$^{1}$Université Paris Cité, CNRS, Laboratoire Matériaux et Phénomènes Quantiques, 75013 Paris, France}
\affiliation{$^{2}$Département de Mathématiques et Applications de l’Ecole Normale Supérieure - PSL, 45 rue d’Ulm, 75230, Paris Cedex 05, France}
\affiliation{$^{3}$ Départment de Physique, Université Paris-Saclay, 91405 Orsay Cedex, France}

\begin{abstract}
Since the early days of quantum information science, diverse strategies have been proposed to encode qubits and implement quantum gates across various physical platforms. Modes, and bosonic states offer a flexible framework for quantum information processing, from single-photon optics to superconducting circuits, combining massless and massive particles. Universal gate sets have been established for specific encodings - such as those based on Fock states or the field's quadratures. However, a general framework clarifying how different bosonic resources - such as mode structure and particle-number statistics - combine to enable universality and potential quantum advantage irrespectively of the encoding remains lacking. A main difficulty lies in the coexistence of distinct paradigms that appear to combine loosely in bosonic quantum information: discrete-variable schemes - where information is encoded in qubits, or effective finite dimensional systems -, and continuous-variable schemes - which rely on infinite-dimensional Hilbert spaces associated with the field's quadratures. Despite the success of each approach, they remain seemingly incompatible in many aspects - in particular in terms of physical resource analysis for universality - due to their underlying mathematical differences. In this work, we introduce a superselection-rule-compliant framework for bosonic quantum information that unifies these paradigms. This representation explicitly incorporates the phase reference and enforces total particle number conservation, allowing for a consistent treatment of bosonic quantum information protocols across both discrete and continuous regimes. We derive necessary and sufficient conditions in terms of bosonic resources for universality that apply to any encoding, without resorting to specific mappings. Our results show that continuous variable encodings are not fundamentally distinct from discrete ones, but rather emerge as subspaces within a general bosonic structure. Finally, we discuss the role of common nonclassicality benchmarks - such as negativity in quasiprobability distributions - in enabling quantum advantage in continuous variables bosonic systems.
\end{abstract}
\pacs{}
\vskip2pc 
 
\maketitle

\section{Introduction}\label{SectionIntro}

Quantum computers are typically built from qubits - or more generally, qudits - and operate within a finite-dimensional Hilbert space. For such systems, various universal gate sets have been identified \cite{PhysRevLett.75.346}. Each gate in these sets acts on a small number of qubits, yet their combinations enable the approximation of any unitary operation on the full $2^n$-dimensional Hilbert space of an $n$-qubit quantum computer, to arbitrary precision. Remarkably, the number of gates required to achieve a given precision scales only polynomially with the system size- a consequence of the Solovay-Kitaev theorem \cite{Kitaev1997,dawson2005solovaykitaevalgorithm, nielsen00, gottesman1998heisenberg}.

Once a computational basis is fixed, gates with different properties should be combined to form a universal set, that potentially enables quantum advantage - that is, the ability to perform tasks that cannot be efficiently simulated by classical computers. A specific subset, known as Clifford gates, can be efficiently simulated classically when acting on computational basis states and followed by measurements in that same basis. This key result is formalized in the Gottesman-Knill theorem \cite{PhysRevA.57.127, PhysRevA.70.052328}.

To be more specific, we define the computational basis of an $n$-qubit quantum computer as formed by the product states $\ket{v_i}$, $i \in \{1,...,2^n\}$
\[
\ket{v_i} = \bigotimes_{k=1}^n \ket{\overline{s_{k(i)}}}_k, \quad s_{k(i)} \in \{0,1\},
\]
where each qubit spans a two-dimensional Hilbert space with the basis $\{\ket{\overline{0}}_k, \ket{\overline{1}}_k\}$. Within this basis, one can define the Pauli group (the Pauli matrices multiplied by the phase factors $\{\pm 1,\pm i\}$), and one commonly used example of universal gate set is given by:

\begin{equation}\label{qubit}
\{\hat H, \hat P, \hat C\} \cup \{\hat T\},
\end{equation}

Here, $\hat H$ is the Hadamard gate, $\hat P=\hat T^2$ and $\hat T$ are different phase gates and $\hat C$ is the controlled-NOT (CNOT) gate. While gates $\hat H$, $\hat P$, and $\hat C$ belong to the Clifford group, $\hat T$ does not. Finally, we define as stabilizer states the image of the computational basis under the application of Clifford gates. The $\hat T$ gate plays then a particular role: when combined with the Clifford gates, it leads to protocols and algorithms executed in a universal quantum computer and that may not be efficiently classically simulable. An alternative to applying the $\hat T$ gate in the circuit model is using \emph{magic states} as resources \cite{PhysRevA.71.022316}.

Magic states are special quantum states obtained by applying the $\hat T$ gate to a stabilizer state given by $\hat H \ket{\overline 0}=\frac{1}{\sqrt{2}}(\ket{\overline 0}+\ket{\overline 1})$:
\[
\ket{T} = \frac{1}{\sqrt{2}}(\ket{\overline{0}} + e^{i\frac{\pi}{4}}\ket{\overline{1}}),
\]

Using $\ket{T}$, the $\hat{T}$ gate can be effectively implemented through gate teleportation protocols \cite{PhysRevA.62.052316}. Moreover, certain noisy states can be distilled into high-fidelity magic states~\cite{PhysRevA.86.052329, Litinski2019magicstate, PhysRevA.95.022316}, and criteria for such distillability have been established in terms of the negativity of phase-space quasiprobability distributions in discrete systems of odd dimension~\cite{Veitch_2012}.

As coherence \cite{RevModPhys.89.041003}, ``magic" (or nonstabilizerness) is relative to a given basis, even though, clearly, not all coherence is magic. Criteria for identifying magic \cite{Wagner_2025} and to relate it to quantum contextuality \cite{ContextMagic} for general discrete encodings have been formalized, in particular, in terms of properties of states' ensembles~\cite{PhysRevLett.126.220404, PhysRevA.109.032220}.

The results discussed above aim to identify the resources that promote efficiently classically simulable systems to universal quantum computing - or that render them not efficiently classically simulable. The goal is then to characterize ``magic" within {\it abstract} computational ({\it i.e.}, mathematical) models, typically with no physical constraints. However, physical systems - although well described by quantum theory - are not, in general, quantum computers. They correspond to Hilbert spaces subject to physical constraints such as symmetries and conservation laws. For example, physical states of bosons are symmetric under particle exchange, which restricts the accessible particle's state space and the set of physically implementable unitaries in multimode bosonic Hilbert spaces. As a consequence, there is an intricate relationship between bosonic modes and particles in information encoding, and in how each contributes as a resource to various quantum information protocols. 

Despite these constraints, quantum information can still be encoded within such systems. One possible strategy is simply to identify two orthogonal physical states to define a computational basis, and then devise methods to implement a universal set of qubit gates for that encoding. Yet, the mathematical structure of the multimode bosonic Hilbert space, combined with the physical properties of the encoding platform, must also be considered. These features introduce specific types of interactions and noise - potentially leading to a variety of error models - that must be described in terms of the encoded information. A simple yet instructive example of the distinction between physical and computational resources is mode coherence. While coherence between modes is a well-defined concept in classical optics - or more generally, in single mode quantum optics, which includes single photons -, it does not play the same role in quantum computation as qubit coherence, as will be further discussed later in this work.

Several encodings with different properties and suited to different physical platforms have been proposed. However, a general result~\cite{Braunstein} identifies a universal gate set for ``continuous variables" (CV), or equivalently, in terms of the quadrature basis defined by
\[
\hat x_k = \frac{1}{\sqrt{2}}(\hat a_k + \hat a_k^\dagger), \quad \hat p_k = \frac{i}{\sqrt{2}}(\hat a_k - \hat a_k^\dagger),
\]
as:
\begin{equation}\label{uni}
\left\{e^{i\hat x_k \alpha},\ e^{i\hat x_k^2 \beta},\ e^{i \hat x_k \otimes \hat x_j \gamma},\ e^{i\frac{\pi}{4} \hat n_k \beta} \right\} \cup \left\{e^{i\hat x_k^3 \zeta} \right\},
\end{equation}
where $k$ and $j$ are arbitrary modes. 
The operations
\[
\left\{e^{i\hat x_k \alpha},\ e^{i\hat x_k^2 \beta},\ e^{i \hat x_k \otimes \hat x_j \gamma},\ e^{i\frac{\pi}{4} \hat n_k \beta} \right\}
\]
are often called quadrature Gaussian (QG) operations, as they preserve Gaussian Wigner functions \cite{PhysRev.40.749}, nonnegative quasi-probability distributions ~\cite{HUDSON1974249, Hudson1}. In contrast, the operation $e^{i\hat x^3 \zeta}$ is quadrature non-Gaussian (QNG) and introduces Wigner function negativities.

Universality in the context of CV quantum computation~\cite{Braunstein} is defined as the ability to approximate, with arbitrary precision, any unitary operation generated by exponentiating a polynomial function of the canonical quadrature operators $\hat{x}$ and $\hat{p}$. Importantly, it has been shown that quantum protocols involving only states, unitaries, and measurements with non-negative Wigner functions in arbitrary representations or basis can be efficiently simulated on a classical computer~\cite{PhysRevLett.109.230503, PhysRevLett.88.097904, Ernesto2}. This result implies that QNG, in particular, and more generally the negativity of  {\it some} CV Wigner function \cite{PhysRevA.102.022411, davis2024}, is a necessary (though not sufficient) condition for achieving quantum computational advantage in CV architectures as well~\cite{Veitch_2014, PhysRevLett.89.207903, Veitch_2012}.

 The physical implementation of a quantum computer is ultimately a physical simulation of an abstract quantum computation \cite{feynman1982simulating, Lloyd:1996aai}. As such, it must consist of protocols or architectures that cannot themselves be efficiently simulated within the physical substrate - in this case, bosonic Hilbert space. It is then not obvious how physical resources (as for instance, QG and QNG operations from \eqref{uni}) should combine so as to efficiently simulate a universal quantum computer for an arbitrary choice of information encoding. From the informational perspective, it is not obvious neither, given some bosonic physical resources, whether protocols using such resources can or cannot be efficiently classically simulated.



To further illustrate these points, we present two representative examples that clarify essential concepts in bosonic information encoding and highlight key open questions - thereby motivating the general framework developed in the following sections. Each example illustrates a distinct encoding strategy in bosonic systems: the first is single-photon encoding, which can be generalized to any fixed-particle-number encoding; the second is CV encoding, in which information - even if discrete- is encoded in an infinite-dimensional Hilbert space where continuous parameters pay an important role, as we will detail in the following.

We start by the single-photon regime, where each qubit is encoded in a pair of orthogonal field modes, with a single photon distributed across them. For instance, the $i$-th qubit of a bosonic quantum computer (BQC) can be defined as:
\[
\ket{0}_i = \hat a^\dagger_{i,0} \ket{\emptyset}, \quad \ket{1}_i = \hat a^\dagger_{i,1} \ket{\emptyset},
\]
where $\ket{\emptyset}$ is the vacuum and the indices $i,0(1)$ refer to orthogonal modes. Modes $i$ might correspond to distinct propagation directions, and $0,1$ might represent polarizations or paths (as in dual-rail encoding~\cite{DualRail}). The annihilation and creation operators obey $[\hat a_{i,0}, \hat a^\dagger_{j,1}] = 0$, $[\hat a_{j,1(0)}, \hat a^\dagger_{j,1(0)}] = \mathbb{1}$ for all $i,j$ and $[\hat a_{i,0}, \hat a^\dagger_{j,0}] = 0$ for $i\neq j$.

To physically implement the universal gate set defined for abstract qubits~\eqref{qubit}, one must perform mode manipulations. Linear optical elements - such as wave plates and time delays - can implement local gates~\cite{KLM, Review2} in the BQC. Such gates are described by passive, QG operations that preserve photon number. While it is possible to define and construct the equivalent to entangling gates using only passive QG operations by encoding quantum information in modes \cite{Scott,PhysRevLett.73.58}, such gates require physical resources (beam-splitters, and modes) that scale exponentially~\cite{Simulation, Simulation2, su, Review2} with the number of qubits, making them not scalable, hence, leading to inefficient simulations of a quantum computer.  Therefore, realizing a two-qubit entangling gate $\hat C$ in a BQC requires nonlinear photon-photon interactions - which turn out to be QNG operations. Notably, this requirement was not initially obvious, as single photons are themselves non-Gaussian in the quadrature representation, and this could have been sufficient for universality, given the established necessary conditions based on the negativity of the Wigner function. As a matter of fact, merely changing the measurement mode basis while restricting to QG gates is sufficient for (strongly believed) quantum advantage, as exemplified by the BosonSampling protocol \cite{v009a004}.

The previous example illustrates that achieving universality in a BQC requires implementing the abstract universal gate set within a concrete physical platform. Since information is encoded in the modes occupied by single photons - defining the computational basis - rather than in abstract qubits, the set of operations enabling efficient exploration of Hilbert space ({\it i.e.}, scalable universality) must necessarily include QNG operations, using \eqref{uni}. In this context, the entangling gate $\hat C$ plays a role analogous to the $\hat T$ gate in the abstract quantum circuit model.

As anticipated, this example reveals a fundamental mismatch between computational and physical resources. It reflects the fact that physical architectures are not abstract quantum computers but rather simulations of them. Even if they successfully reproduce the abstract gate operations required for universality, they do so within a Hilbert space governed by distinct physical constraints and encoding strategies. 

It is now important to turn to another widely used quantum information encoding strategy in bosonic systems, that consists of encoding a qubit in orthogonal states of a single mode. Examples of such approaches, that we will call from now one the CV encoding, include cat codes~\cite{MazyarParadigm, PhysRevA.87.042315} and the Gottesman-Kitaev-Preskill (GKP) encoding~\cite{gottesman_encoding_2001}. In these schemes, logical qubits are defined using specific superpositions within a single mode, leading to:
\[
\ket{\overline{0}}_i = \sum_{n=0}^\infty c_n^{(0)} \ket{n}_i, \quad 
\ket{\overline{1}}_i = \sum_{n=0}^\infty c_n^{(1)} \ket{n}_i, \quad 
{}_i\langle{\overline{0} | \overline{1}}\rangle_i = 0,
\]

where mode $i$, as before, encodes the $i$-th qubit. The specific states used in this type of encoding are often chosen for reasons such as their robustness to certain types of noise that operate in the infinite dimensional CV Hilbert space or for experimental feasibility (e.g., squeezed states~\cite{PhysRevLett.112.120505, PhysRevA.109.040101, PhysRevLett.114.050501}). For each encoding, an analog of the gate set~\eqref{qubit} can be defined, typically involving both QG and QNG operations from~\eqref{uni}. However, as in the case of single photons, the necessary and sufficient bosonic requirements for universality irrespectively of the encoding - such as the exact combination of QG and QNG resources - remain unclear, as whether or how these requirements depend on the chosen encoding.

Unlike the single-photon encoding, the CV encoding makes explicit use of the continuous nature of quadrature representations and the infinite dimensionality of single-mode Hilbert spaces. These aspects may themselves be considered as either resources or limitations. While infinite-dimensional spaces are often associated with systems of infinite energy - and those are clearly unphysical - in practice, CV encodings rely on finite-energy approximations of ideal states (as for instance, quadrature eigenstates). Errors are then intrinsic to the CV encoding. For these reasons, comparing or mapping these computational spaces to the finite-dimensional Hilbert spaces of discrete-variable quantum computing is not straightforward, and some attempts combining both discrete and continuous aspects of states and operations can be found, for instance, in \cite{calcluth2023sufficient, PhysRevA.104.012430, ModularMenicucci,Pierre, Andreas}. This operational difficulty poses additional challenges and strategies for solving them when attempting to transpose known results from discrete variables quantum computing to the CV encoding~\cite{arzani2025effectivedescriptionsbosonicsystems, Veitch_2012, PhysRevLett.126.190504, Marshall:23}.

We have thus briefly analyzed two examples of encoding strategies, each apparently relying on different physical features of bosonic systems. In the single-photon encoding, different qubit states are associated with the same field state - a single photon - distributed across different modes, thereby defining a finite-dimensional Hilbert space. In contrast, CV encoding uses orthogonal states within a single mode to represent qubits, and the infinite-dimensional, continuous nature of the field in each mode plays a central role. This duality in the nature of encodings - whether using different modes for a fixed photon number, or different states within the same mode - already suggests the need for a deeper understanding of the principles governing information encoding in arbitrary field states and bosonic systems in general.

Moreover, we have seen that the role of computational resources (as different gates, or merely the meaning of coherence) for the general results developed for abstract qubits is not the same for bosonic systems, at least for the studied encodings. This leads to open questions about the role of various bosonic resources - such as modes, states, QG and QNG operations - for quantum information. While these issues have been investigated to some extent by the community, they remain unsolved, and their connection to the fundamental problem of the relation between different bosonic encodings has not been addressed. This raises a central question: how should gates from the set~\eqref{uni} be combined to efficiently simulate a universal quantum computer under an arbitrary encoding? More specifically, what are the necessary and sufficient resources - quantified in terms of, for instance, QG and QNG operations - for achieving universality? And what is the minimal amount of QNG, or any other resource, or combination of resources, required to enable potential quantum advantage? 

In this contribution, we address these questions and make significant progress toward providing general answers by adopting a new perspective on information encoding and manipulation in bosonic systems. The remaining of the paper is organized as follows: in Section \ref{SectionReview}, we review and discuss prior results that have explored these open questions, focusing primarily on the CV formalism that presents more open questions for being an intrinsically infinite dimensional system with, potentially, infinite energy. Section \ref{SectionMain} summarizes our main contributions. Section \ref{SectionSSRC} introduces in detail the formal framework we employ, based on a superselection rule compliant (SSRC) formalism, which allows us to rigorously address and resolve the challenges identified. In particular, we analyze how the CV representation emerges as a particular case of our formalism and the consequences of adopting such particular case. Our results are then presented in Section \ref{SectionGeneral} and Section \ref{SectionDiscussion} contains the discussion of these results and concluding remarks.

\section{Brief review and discussion of previous results}\label{SectionReview}



Before presenting the main contributions of this work, we briefly review key developments in quantum information encoding and the identification of quantum resources in bosonic systems.

In Ref. \cite{MattiaX}, an overview of QNG states is provided, emphasizing in particular the necessity of QNG entanglement for achieving quantum computational advantage - a point also identified in \cite{PhysRevLett.130.090602}. 

The limitations of necessary conditions alone for non efficient classical simulability in bosonic systems have been highlighted by several works (for example, \cite{PhysRevResearch.3.033018, Calcluth2022efficientsimulation}): circuits or protocols may involve highly QNG states (as quantified, for instance, by QNG quantifiers, as the stellar rank \cite{Stellar}) yet can still be efficiently classically simulated. A more complete framework is proposed in \cite{calcluth2023sufficient}, where necessary and sufficient conditions for quantum universality are established in terms of (ideal) GKP encoding, generalizing important achievements on the continuous-to-discrete mapping of quantum information \cite{PhysRevA.104.012430, ModularMenicucci,Pierre, Andreas}. In short, in \cite{calcluth2023sufficient} a hybrid architecture is constructed by encoding information in highly QNG states - specifically, ideal GKP states - and use operations that, when combined with this encoding, lead to circuits that can be efficiently classically simulated. Different states can complete this architecture and the authors propose a resource-based classification of such CV states, based on the magic they provide within this computational framework. In \cite{davis2024}, a related result is obtained from the phase-space perspective. The authors show that GKP states possess a non-negative phase-space representation under a well-chosen quasi-probability distribution \cite{Hudson1}. Hence,``magic" for this particular encoding is tied to the negativity of the chosen phase-space representation. These frameworks are consistent with previous results showing that states that are QG, such as coherent states, can promote GKP-based circuits to universality~\cite{NGaussianMenicucci, PhysRevA.107.062414}.

These results suggest that the capacity of a state or operation to promote an architecture to universality - that is, to inject ``magic" into a given bosonic computational model - is a relative and collective property, rather than an absolute one. Just as abstract qubits, or qudits are not inherently resourceful or resource-less - unless a specific computational basis and set of operations or states is specified - absolute statements cannot be made about specific bosonic states, operations or resources in general. It is clear that, when analyzing physical systems, one may be more tempted to associate physical properties to absolute constraints, partially due to the difficulties related to their implementability across different platforms and to the notion of ``classicality". However, as shown for instance in \cite{PhysRevD.24.1516}, even this notion is relative to a pointer basis. Nevertheless, it appears from the previously studied examples, that both QG and QNG must appear as ingredients to construct logical gates in bosonic encodings.


Ref.~\cite{descamps2024superselectionrules} performs a paradigm change by identifying a key limitation of the CV formalism with important consequences on the applications of bosonic systems in quantum information: the lack of an explicit treatment of phase references and, consequently, of superselection rules (SSR)s. By introducing a formulation of bosonic states that explicitly respects particle number conservation and treats the phase reference as a physical resource, the authors develop a mapping from arbitrary bosonic states with explicit quantum phase references to well-defined states in a BQC. Within this mapping, which extracts the first-quantization properties of the system and transform them into second quantization ones \cite{PlenioExtract, PhysRevX.10.041012}, bosonic states prepared via particular QNG operations (that will be specified later in this contribution) correspond to BQC states involving the entangling gate $\hat C$
 - a known ``magic" resource in the BQC framework (which is also a QNG operation). While this is not, at this point, immediately obvious (see Section \ref{SectionSSRC}), the proposed mapping provides a unified physical and computational criterion for nonclassicality that is valid both in the CV limit and for states with a finite number of photons. In particular, genuinely multimode states (which cannot be reduced to single-mode states via passive linear optics) are systematically identified as nonclassical.

A key contribution of \cite{descamps2024superselectionrules} is the adoption of a SSRC formalism \cite{PhysRevA.68.042329, PhysRevA.55.3195, PhysRevA.58.4244,PhysRevA.58.4247,doi:10.1142/S0219749906001591,Sanders_2012,PhysRev.155.1428, doi:10.1142/S0219749906001591, PhysRevD.27.2885}, which reveals that the standard CV representation can be understood as a limiting case of a more general representation in an arbitrarily large but finite-dimensional Hilbert space. This enables, in principle, differently from the mapping proposed in ~\cite{descamps2024superselectionrules}, a mapping between bosonic states and abstract finite-dimensional (qubits and qudits) spaces, where criteria for universality are well established for abstract quantum computers. However, such mappings do not permit identifying  the sufficient physical resources for universality given an arbitrary encoding in a bosonic system, since they disregard the physical characteristics of the original bosonic system. It turns out that the main focus of the present work is precisely determining how physical resources contribute to computational resources. In other words, simply mapping a bosonic simulation of a quantum computer into abstract qubits cannot directly inform us about the necessary and sufficient {\it physical} conditions for efficient simulation of a quantum computer and of specific gates using the original bosonic architecture. For instance, it occurs, as observed in the single photons encoding, that Clifford operations in an abstract quantum computer corresponds to QNG ones in a BQC.  

Finally, in Ref.~\cite{descamps2024superselectionrules}, a privileged basis is introduced based on particle separability - a physically motivated criterion that will be further discussed in the following sections. This choice of basis allows one to define and identify informational resources relative to it. While various bosonic encodings can be mapped into different BQC encoding schemes (including those based on entangled states), it is not evident that identifying the physical resources required for universality for arbitrary encodings becomes simpler or more transparent using the BQC mapping. For this reason, and in order to maintain a direct connection with the most commonly used bosonic encodings, especially in CV, we have not adopted that mapping in the present work.

\section{Main results}\label{SectionMain}
 
Our main contribution is to explore the SSRC as a new framework for encoding and processing information in bosonic systems. This consists of a complete change of perspective that brings significant consequences for bosonic quantum information, as follows. 

First, the SSRC framework enables a rigorous identification of the necessary and sufficient physical resources required for universal quantum computation with bosonic systems - independently of the specific encoding. By associating an arbitrary bosonic state to a qubit's state and analyzing operations within the quantum circuit model we show how, for any encoding, the physical resources required to construct a universal gate set as \eqref{qubit} are linked to QG and QNG operations. Surprisingly, we find that these resources are encoding-independent, with the only exception being the single-photon encoding.

Second, we demonstrate that the CV representation corresponds to a specific limit of the SSRC framework. By examining this limit, we uncover subtle and often overlooked aspects of CV quantum information processing. Notably, we show some counterintuitive results regarding the representation dependency of the  notions of QG and QNG operations; first, they depend on the choice of a {\it mode} basis. Second, these notions emerge from approximations applied to bosonic states that restrict the accessible Hilbert space, rather than reflecting fundamental properties, and such approximations are basis and state dependent. This relativity suggests that the CV framework—whether in its quadrature form, modular form \cite{Aharonov:1969qx}, or other variants —does not provide the most transparent or fundamental description of resources for bosonic quantum computation. As a consequence, our results suggest that widely used indicators of non-classicality - such as CV Wigner functions negativity, stellar rank, or other phase-space-based metrics - are representation-dependent. Consequently, they do not capture the full structure of quantum resources across encodings. In contrast, our results establish the general necessary and sufficient conditions for universality that are valid across all bosonic encodings, and not just in the CV regime. 

Third, our formalism unifies a wide range of bosonic information encoding schemes - such as single-photon \cite{DualRail, KLM}, binomial \cite{PhysRevX.6.031006}, cat \cite{MazyarParadigm, PhysRevA.87.042315}, GKP~\cite{gottesman_encoding_2001} and other CV encodings - into a single coherent theoretical structure. This unified picture clarifies the roles of modes and states, and how they are manipulated, offering a consistent language to describe encoding, transformations, and measurement in bosonic systems.

Last but not least, the SSRC representation allows one to provide a physical interpretation to different gates of the universal gate set. In particular, the presence or absence of particle interactions enables characterizing the different bosonic gates - a connection that is absent in models based on Eq.~\eqref{uni}. This interpretation provides a physically grounded understanding of the relevant combination of resources - as modal and particle properties - for potential classical non-simulability.

Finally, from a physical standpoint, we reveal that the mode-dependence of CV states can be traced back to the choice of vacuum. This understanding allows us to resolve the physical meaning (and limitations) of common tools, such as quadrature eigenstates, which are known to be non-physical. Our framework thereby supports more consistent and physically grounded definitions of key concepts such as coherence, entanglement, and noise in bosonic systems.

\section{The superselection rule compliant representation}\label{SectionSSRC}

We now summarize the SSRC representation and discuss some of its properties and associated tools \cite{PhysRevA.68.042329, PhysRevA.55.3195, PhysRevA.58.4244,PhysRevA.58.4247,doi:10.1142/S0219749906001591,Sanders_2012,PhysRev.155.1428, doi:10.1142/S0219749906001591, PhysRevD.27.2885}.

The conventional description of single-mode states of the electromagnetic field - commonly referred to as the CV representation, as discussed in Section \ref{SectionIntro} - relies on superpositions of Fock states. In this representation, a single mode state is represented as $\ket{\psi}_a = \sum_{n=0}^{\infty} c_n \ket{n}_a$, where the subscript $a$ denotes the considered mode and the normalization condition $\sum_{n=0}^{\infty} |c_n|^2 = 1$ holds. This standard formulation implicitly assumes the existence of a phase reference: otherwise, the state $\ket{\psi}_a$ violates the photon number SSR. If no phase reference is specified, $\ket{\psi}_a$ would instead have to be replaced by a statistical mixture $\hat \rho = \sum_{n=0}^{\infty} |c_n|^2\ketbra{n}$, satisfying $[\hat \rho, \hat n_a] = 0$, with $\hat n_a=\hat a^{\dagger}\hat a$, or by a pure state with a well-defined total photon number (see Eq. \eqref{Psi} below). In short, the CV representation is only operationally valid when a phase reference is explicitly or implicitly defined.

It is possible to represent pure states that are SSRC by explicitly incorporating the phase reference, which is an example of a quantum reference frame \cite{PhysRevD.95.043510,PhysRevD.27.2885,PhysRevD.30.368,giovannetti_quantum_2015,QRF}. In this case, the phase reference is quantized and is associated with a Hilbert space. It can be considered as a physical resource - as it is, even when implicit \cite{PhysRevX.10.041012,RevModPhys.89.041003, descamps2024superselectionrules}. This approach avoids relying on implicit assumptions and provides a consistent treatment of states and operations in quantum optics and bosonic systems in general. A  discussion of some of the implications of using the SSRC representation for quantum information can be found in ~\cite{RevModPhys.79.555, descamps2024superselectionrules, Saharyan2025}.

Remarkably, the SSRC representation permits unifying the single-mode CV regime with multimode entangled states with fixed total photon number under a common formal structure, as we now show. For this, we can consider the simplest SSRC pure state, that contains an arbitrary but fixed number of photons $N$:
\begin{equation} \label{Psi}
\ket{\Psi} = \sum_{n=0}^{N} c_n \ket{n}_a \ket{N - n}_b,
\end{equation}
where modes $a$ and $b$ are orthogonal and serve as mutual phase references (as for instance, two spatial modes), $\sum_{n=0}^N |c_n|^2=1$, $\ket{N}_{a(b)}= \left (\hat a^{\dagger} (b^{\dagger})\right )^N/\sqrt{N!}\ket{\emptyset}$, and $\ket{\emptyset}$ is the vacuum state. We notice that  \eqref{Psi} has $U(1)$ symmetry and is invariant under the transformation $e^{i\phi(\hat a^{\dagger}\hat a+\hat b^{\dagger}\hat b)}$. The CV regime can be recovered as a limiting case where most of the photons occupy a single mode (here, mode $b$), {\it i.e.}, $\sum_{n=0}^{N} |c_n|^2 n \ll N/2$. We can then formally consider $N \to \infty$. In this limit, it is also always possible to  
define a formal cutoff $n_{\rm max}$ such that $N \gg n_{\rm max}$, with $n_{\rm max} \gg \sum_{n=0}^{n_{\rm max}} n |c_n|^2$. In this case, we can represent $\sum_{n=0}^{N} c_n\ket{n}_a \ket{N-n}_b \to \sum_{n=0}^{n_{\rm max}} c_n\ket{n}_a$, omitting the phase reference (mode $b$)  \cite{PhysRevA.68.042329, PhysRevA.55.3195, PhysRevA.58.4244,PhysRevA.58.4247,doi:10.1142/S0219749906001591,Sanders_2012,PhysRev.155.1428, doi:10.1142/S0219749906001591,RevModPhys.79.555} and let $n_{\rm max} (\ll N) \to \infty$, so $\ket{\Psi} = \sum_{n=0}^{N} c_n \ket{n}_a \ket{N - n}_b \to \sum_{n=0}^{\infty} c_n\ket{n}_a$, that we will call from now on {\it the CV limit}. The procedure described explicits that the measure of the photon number imbalance between modes is relative to the average energy of mode $a$. Since physical states have finite energy, any single mode CV state can be seen as a SSRC state in the form \eqref{Psi} with a large photon number imbalance between modes. 

However, in some situations, the CV representation is less suited. This is the case of experiments with a fixed number of photons distributed over many modes, as the Hong-Ou-Mandel experiment  \cite{PhysRevLett.59.2044}, where different occupied modes serve as phase references to one other. We can also notice that the CV representation is not necessary to observe coherence and interference phenomena even when the phase reference cannot be directly measured and states in mode $a$ should be represented as statistical mixtures (obtained by tracing out mode $b$ in \eqref{Psi}) \cite{PhysRevA.68.042329, PhysRevA.55.3195, PhysRevA.58.4244,PhysRevA.58.4247,doi:10.1142/S0219749906001591,Sanders_2012,PhysRev.155.1428, doi:10.1142/S0219749906001591}: in this case, coherence can be created by mode manipulations at the measurement stage \cite{MeasurementInduced, PhysRevA.55.4330,PeterHorak1999}. 

It is important to stress that no approximation is made in writing Eq.~\eqref{Psi}. Rather, it is the single-mode CV representation that emerges as a particular case - valid when one mode's population dominates. Hence, \eqref{Psi} remains a valid representation of bosonic states in any case, allowing for consistent treatment both of states that reduce to CV-like ones under large photon-number imbalance between modes (as previously shown), and those where the two mode's populations are balanced. This observation has important consequences, especially considering that a strategy to overcome the mathematical and physical difficulties associated with quantum information encoding with CV is imposing physical constraints, as for instance in energy  \cite{PhysRevLett.126.190504}. Such constraints are not in fact necessary in the present formalism, since it is the CV limit that is shown to be a sometimes convenient description, but it comes with the price of introducing an apparently different mathematical formalization of bosonic systems that hides some of its important physical and mathematical properties. It is of course also possible to describe CV multimode states, {\it i.e.}, multimode states with a non-fixed photon number, using the same type of limit, as we will discuss later. Thus, the SSRC representation offers a general framework that preserves physical consistency across different regimes and provides a more complete view of the structure of quantum resources in bosonic systems for explicitly including the phase reference.

Another interesting feature of the representation used in Eq. \eqref{Psi} is that it is also used to represent states of a $j=N/2$ angular momentum system and symmetric states of $N$ two-level systems  \cite{Schwinger, Sakurai}. Hence, the SSRC representation unifies the treatment of diverse bosonic systems and symmetric states \cite{RMPSym} - such as cold atoms, Dicke states, mechanical resonators and photons  - under a single theoretical framework. Alternatively, symmetric states can be expressed in a form due to E. Majorana \cite{Majorana1932AtomiOI, Vieta, PhysRevA.87.053821} that is extremely useful \cite{Giraud_2010, Extremal,PhysRevResearch.5.L032006, Multipoles, PhysRevA.92.031801,Aaron}:
\be\label{Majorana}
\ket{\psi}= \frac{1}{\cal N}\prod_{k=1}^N(\cos{\left (\frac{\theta_k}{2}\right )}\hat a^{\dagger}+e^{i\phi_k}\sin{\left (\frac{\theta_k}{2}\right )}\hat b^{\dagger})\ket{\emptyset},
\ee
where ${\cal N}$ is a normalization constant, and $\theta_k \in [0,\pi]$, $\phi_k \in [0,2\pi]$.  

In any form, \eqref{Psi} or \eqref{Majorana}, using the SSRC representation enables the direct application of results from \cite{PhysRevLett.132.153601}, which identified a universal gate set for symmetric subspaces of $N$ spin $1/2$-like systems. We can define angular momentum operators $\hat J_x^{(i,j)}=(\hat a^{\dagger}_i \hat a_j + \hat a_i \hat a^{\dagger}_j)/2$, $\hat J_y^{(i,j)}=i(\hat a_i\hat a^{\dagger}_j-\hat a^{\dagger}_i \hat a_j)/2$, $\hat J_z^{(i,j)}=(\hat a^{\dagger}_i\hat a_i - \hat a^{\dagger}_j\hat a_j)/2$, where $N= n_i + n_j$ is the total number of photons in two arbitrary orthogonal modes $i, j$. Using these operators it is possible to construct linear $e^{i\chi \hat J_{\vec{n}}^{(i,j)}}$ and non linear $e^{i\chi (\hat J_{\vec{n}}^{(i,j)})^2}$ SSRC unitary operations, where $\vec{n}$ is a unit vector, the projection direction of the angular momentum. These operations form a universal set, capable of transforming any two states of the form \eqref{Psi} into one another with arbitrary precision \cite{PhysRevLett.75.346, doi:10.1098/rspa.1995.0065}, using a number of operations scaling at most polynomially in $N$. Finally, these results can be extended to a multimode setting, where a general $K$-mode SSRC pure state is given by $\ket{\Psi}=\sum_{\{n_i\} : \sum_i^K n_i= N}^N c_{n_1,...,n_K}\ket{n_1}_1...\ket{n_K}_K$. In this case, a multimode unitary gate set can be built with the operations $e^{i\chi \hat J_{\vec{n}}^{(i,j)}}$ and $e^{i\chi (\hat J_{\vec{n}}^{(i,j)})^2}$, where $i,j ~\in ~ \{1,...,K\}$ are arbitrary orthogonal modes (see details in Appendix \ref{App1}).

The unitary operators $e^{i\chi \hat J_{\vec{n}}^{(i,j)}}$ have different physical interpretations according to the physical system at hand: they represent rotations of angular momentum states and they are passive QG operations. Generally, they are linear SSRC operations that perform a mode basis change, and we will refer to them in the following as SG (short for superselection rule compliant Gaussian) operations. As for unitaries involving quadratic and higher powers of  $\hat J_{\vec{n}}^{(i,j)}$, they represent a spin squeezing-like interaction, and they are SSRC gates corresponding to QNG operations. They will be noted here SNG (short for superselection rule compliant non-Gaussian) . The present construction perfectly mirrors the results in \cite{Braunstein}, which identified a universal gate set for the CV representation, Eq. \eqref{uni}. Nevertheless, as the CV representation is a limit of the SSRC one, we expect that \eqref{uni} can also be obtained from the SSRC universal gates. We will indirectly show this later in this contribution. For the moment, it is interesting to stress that, contrary to \eqref{uni}, different SSRC universal gates have fundamentally distinct physical properties from one another. While the linear operations $e^{i\chi \hat J_{\vec{n}}^{(i,j)}}$ are mode basis changes that act {\it locally} on each particle's state if one adopts the first quantization formalism, the unitaries $e^{i\chi (\hat J_{\vec{n}}^{(i,j)})^2}$ are associated with particle {\it interactions} in first quantization. This physical distinction between the QG and the QNG operations is less clear, since a QG unitary may describe particle interactions (as the shearing operator $e^{i\alpha \hat x^2}$), or merely displace states in phase space without modifying them. As announced, the origin of such a physical inconsistency in the quadrature representation will become clearer in the discussions that follow.  

Another interesting distinction between the SG and SNG concerns how they manipulate different properties of an arbitrary state. Inspecting \eqref{Majorana}, we see that state with a total photon number $N$ is characterized by $2N$ (different or not) real numbers, ($\theta_i$, $\phi_i$), $i=1,...,N$. While SG operations do not change the degeneracy of these values, the SNG ones do, a point that is directly related with the study of roots of the Majorana polynomial that is out of the scope of the present contribution. An alternative way to present this result using the quantum optics vocabulary is by stating that SG operations are simply mode basis changes, that preserve the {\it intrinsic} properties of the states, implementing the transformation $e^{i\chi \hat J_{\vec{n}}^{(i,j)}}\ket{\psi}_{a,b} \to \ket{\psi}_{c,d}$, where $c$ and $d$ are orthogonal modes obtained from modes $a$ and $b$ by a rotation around $\vec{n}$ by an angle $\chi$, while SNG operators affect the {\it state} in any mode basis $\ket{\psi}=\sum_{n=0}^N c_n\ket{n}_a\ket{N-n}_b \to \sum_{n=0}^N d_n\ket{n}_{c}\ket{N-n}_{d}$, with $\{c_n\} \neq \{d_n\}~\forall~c,d$. 

We will discuss some important consequences of these facts when analyzing specific examples in the next Section.

s
\subsection{The continuous limit and the relativity of quadrature non-Gaussianity}

In this Section, in order to simplify the notation, we will omit the superscripts $(i,j)$, since we will mainly focus on states in the form \eqref{Psi}, \eqref{Majorana}, where the mode $a$ will correspond to the considered single mode in the CV limit and the mode $b$ will play the role of the explicit phase reference.

As we have seen, every state in the CV representation has an equivalent formulation in the SSRC representation, which offers the advantage of clearly distinguishing between particle-dependent (intrinsic) and mode-dependent (relative) properties of bosonic states as we started to see and will continue to analyze in the following. For this, we now see in more details how states and operations in the CV representation emerge from the SSRC framework in the limit $n_{\rm max}~ (\ll N) \to \infty$, using representative states to illustrate the principle. In particular, we will focus on showing how the CV representation may obscure relevant physical properties of bosonic states. We start from Fock states, such as $\ket{N}_b$, that have been identified as SSRC classical-like states in prior works \cite{descamps2024superselectionrules, PhysRevX.10.041012, PhysRevA.96.032304, HP2}. The reason for this identification is that Fock states are separable particle states (in the first quantization representation). Fock states also correspond to the so called spin coherent states \cite{PhysRevA.78.042112, Byrnes_Ilo-Okeke_2021, perelomov_generalized_1977, JMRadcliffe_1971, PhysRevA.6.2211}.

 In the CV representation, one associates classical-like states to Glauber's coherent states \cite{PhysRev.131.2766}, $\ket{\alpha}_a=\sum_{n=0}^{\infty} \frac{\alpha^n}{\sqrt{n!}}e^{-\frac{|\alpha|^2}{2}}\ket{n}_a$, where $a$ is a given mode. In coherent states, particles (photons) behave independently from one another, as for Fock states. Nevertheless, there are, apparently, some important differences between these states: coherent states do not have a fixed particle number, they have a well defined phase, and they are considered as QG states, while Fock states are QNG states. However, using the SSRC representation in different limits, it is possible to show that Fock states and coherent states are intrinsically the same, and that the CV representation omits several interesting physical aspects of the coherent states.

We start by inspecting \eqref{Majorana}, that shows that Fock states are fully characterized by only two real parameters (one complex number) for a fixed $N$ ($\theta_k=\theta$,$\phi_k=\phi$, $\forall~k$, with respect to the reference mode - say, $b$ - in \eqref{Majorana}). Also, classical fields and Glauber's coherent states are fully characterized by a complex number, but $N$ does not seem to play a role for such states. Nevertheless, we can notice that a general Fock state $\ket{N}_{\theta,\phi}$ is obtained by the applications of a SG operation (rotation) on $\ket{N}_b$, that can be expressed as $e^{i\phi \hat J_z} e^{i\theta \hat J_y}= \hat R(\theta, \phi) (= e^{i\chi \hat J_{\vec{n}}})$, with $\hat R(\theta, \phi)\ket{N}_b=\ket{N}_{\theta,\phi}= \frac{1}{\sqrt{N!}}(\hat c_{\theta,\phi}^{\dagger})^N\ket{\emptyset}$, and $\hat c_{\theta,\phi}^{\dagger}=\cos{\left ( \frac{\theta}{2}\right )}\hat b^{\dagger}+e^{i\phi}\sin{\left ( \frac{\theta}{2}\right )}\hat a^{\dagger}$. When $a$ and $b$ are propagation modes, $\hat R(\theta, \phi)$ is implemented with a combination of beam splitters and differences in optical path lengths.

Using such transformations, we see that the overlap between two Fock states ${}_{b}\langle N|N\rangle_{\theta,\phi}=(\cos{[\theta/2]})^N$ depends on the {\it mode} overlap $[\hat c_{\theta,\phi}, \hat b^{\dagger}] = \cos{[\theta/2]}$ \cite{TrepsModes}. The $N$ dependency of the state's overlap enhances the (classical-like) phase locking of two Fock states: for $N \gg 1$ \cite{PhysRevA.68.042329, MeasurementInduced}, $|{}_{b}\langle N|N\rangle_{\theta,\phi}| \rightarrow e ^{- \frac{N\theta^2}{8}}$, approaching a delta function centered at $\theta=0$: $N$ exponentially amplifies the effect of a non-perfect mode overlap. This suggests that classical states can be identified with the $N \to \infty$ limit of Fock states.

However, it is possible to provide a more refined picture of the phase locking process by defining a $N$ independent parameter $\alpha$ such that $\ket{N}_{\alpha, N}=\sqrt{N!}^{-1}(\hat c^{\dagger}_{\alpha,N})^N\ket{\emptyset}$, with $\hat c^{\dagger}_{\alpha,N}=\hat R((\theta,\phi)(\alpha,N) )\hat b^{\dagger}\hat R^{\dagger}((\theta,\phi)(\alpha,N) )= \sqrt{1-\frac{|\alpha|^2}{N}}\hat b^{\dagger} + \frac{\alpha}{\sqrt{N}}\hat a^{\dagger}$,(see Appendix \ref{App2} for details). This construction generalizes for arbitrary $N$, and for two complex amplitudes $\alpha,\beta$, ${}_{\beta,N}\langle N|N\rangle_{\alpha,N}=\left (\sqrt{1-\frac{|\alpha|^2}{N}}\sqrt{1-\frac{|\beta|^2}{N}}+\frac{\alpha\beta^*}{N}\right )^N$.  Modes and states are orthogonal iff ${\rm arg}[\beta] ={\rm arg}[\alpha]+\pi$, and $|\alpha|^2+|\beta|^2=N$.  However, it is possible to study how the state's overlap approaches zero. For $N \gg |\alpha|^2, |\beta|^2$, $|{}_{\alpha,N}\langle N|N\rangle_{\beta,N}| \simeq e^{-\frac{1}{2}|\alpha-\beta|^2} = |\langle \beta|\alpha \rangle|$, where $\ket{\alpha(\beta)}$ are Glauber's coherent states, providing a more refined picture of the phase locking process and of the relation between {\it classical} SSRC states and the {\it coherent} CV ones. In particular, the latter form a subspace of classical SSRC states characterized by small rotation angles ($\theta \approx 2|\alpha|/\sqrt{N} \ll 1$) with respect to the phase reference mode (mode $b$ in the example).
Interestingly, while coherent CV states in mode $a$ appear as single mode Fock states superpositions, with amplitudes that are operationally independent of the phase reference mode $b$, their properties, given by the complex amplitudes of each Fock state, actually implicitly relies on the coherence between these modes. As a matter of fact, the {\it modal} properties of SSRC classical states are transferred into {\it statistical} properties and coherence between different Fock states in the CV representation. In particular, we can notice that the definition of the vacuum state itself is a relative one, and it is, not surprisingly, not associated with the absence of energy. In CV, the vacuum corresponds to state $\ket{N}_b$ in the present mode basis choice ($\theta=0$, $|\alpha|=0$), which is a classical state in the phase reference mode. Consequently, intrinsically, all coherent states are Fock states in different modes or, alternatively, they each correspond to different reference modes (in the example, we have chosen mode $b$ as the phase reference in the CV limit), or different choices of the vacuum state.

An alternative way to see this, is using that $\hat R((\theta,\phi)(\alpha,N) )\ket{N}_b=\ket{N}_{\alpha,N}$. Hence, {\it displacements of coherent states in phase space} - implemented, in the CV representation, by the unitary operators $\hat D(\alpha)= e^{\alpha \hat a^{\dagger}-\alpha^*\hat a}$, with $\hat D(\alpha) \ket{\emptyset}=\ket{\alpha}$  - correspond to {\it mode rotations} of classical states in the SSRC representation. In the Appendix \ref{App3} we show that the mapping between mode rotations and phase space displacements can always be done if $n_{\rm max} \ll N$, which is always satisfied in the CV limit. This is an example of how an operation of the universal gate set in the quadrature basis \eqref{uni} appears as a limit of a SSRC operation. 

In the previous example, a SG operation appears as a QG one in the CV limit. However, if we analyze states, we see that Fock states, in spite of being classical, are QNG states.  Nevertheless, they appear as QG in the CV limit. This indicates that QNG quantifiers are not consistent with the intrinsic physical properties of bosonic states and operations. 
 
We now examine another instructive example, by following similar procedures as above. The state $\ket{N}_{\theta_1,\phi_1}\ket{N}_b=\frac{1}{{\cal N}}(\cos{(\theta_1/2)}\hat a^{\dagger}+e^{i\phi_1}\sin{(\theta_1/2)}\hat b^{\dagger})^N(\hat b^{\dagger})^N\ket{\emptyset}$ is characterized by four parameters (for fixed $N$  \cite{NoteS}) in \eqref{Majorana}: $\theta_{1(2)}, \phi_{1(2)}$, relative to a reference mode basis ($a$ and $b$, so $\theta_{1\leq k \leq N}= \theta_1$, $\phi_{1\leq k \leq N}=\phi_1$ and $\theta_{ N+1 \leq k \leq 2N}=\theta_2=\pi$, $\phi_{ N+1 \leq k \leq 2N}=\phi_2=0$).

We can choose to express $\ket{N}_{\theta_1,\phi_1}\ket{N}_b$ in a rotated basis $\ket{\cal S}=\hat R((\eta,\chi)(r,\varphi))\ket{N}_{\theta_1,\phi_1}\ket{N}_b \propto (\hat c^{\dagger})^N(\hat d^{\dagger})^N\ket{\emptyset}$, with $\hat c= e^{-\frac{r}{2}}(e^{i\frac{\varphi}{2}}\sqrt{\cosh{r}}\hat b + \sqrt{\sinh{r}}\hat a)$, $\hat d= e^{-\frac{r}{2}}(-e^{i\frac{\varphi}{2}}\sqrt{\cosh{r}}\hat b + \sqrt{\sinh{r}}\hat a)$. Then, up to a normalization constant, 
 \be\label{squeezie}
\ket{\cal S} \to \sum_{n=0}^N \binom{N}{n}(-\tanh{r})^{n}e^{i\varphi (N-n)}(\hat a^{\dagger})^{2n}(\hat b^{\dagger})^{2(N-n)}\ket{\emptyset}.
 \ee
As we have seen, the mode basis choice corresponds to a phase-space displacement or a modification of the phase-space origin, which does not alter the intrinsic physical properties of the state. Equivalently, the parameters $\theta_k, \phi_k$ in \eqref{Majorana} are invariant under a mode basis changes, and they can be used to identify different families of CV states from a SSRC perspective. A relevant family of CV states that is defined by four parameters, as state $\ket{N}_{\theta_1,\phi_1}\ket{N}_b$, are squeezed states $\ket{r, \varphi}_{\alpha} =\hat S(\zeta)\ket{\alpha}$, where $\hat S(\zeta)=e^{\frac{1}{2}\left (\zeta (\hat a^{\dagger})^2-\zeta^*(\hat a)^2\right )}$ is the squeezing operator, with $\zeta = r e^{i\varphi}$, $r$ the squeezing  parameter, $\varphi$ its phase and $\alpha$ is the amplitude of an initial coherent state \cite{NoteS2}. If $\tanh{r} < 1$ (finite squeezing, physical states) and $N \rightarrow \infty$, we can always define a cut-off $ n_{\rm max}$ sufficiently large, with $1 \ll n_{\rm max}\ll N$ so that, up to a normalization factor, Eq. \eqref{squeezie} becomes 
\be\label{nice}
\ket{r,\varphi}_0\propto \sum_{n=0}^{n_{\rm max}}\frac{(-e^{-i\varphi }\tanh{r})^{n}\sqrt{(2n)!}}{2^{n}n!}\ket{2n}_a\ket{2(N-n)}_b. 
\ee
so  $\ket{N}_{\theta_1,\phi_1}\ket{N}_b \to \ket{r,\varphi}_0$ (details in Appendix \ref{App4}).

Eq. \eqref{nice} corresponds to the squeezed vacuum state, characterized by $\alpha=0$, $r$ and $\varphi$ or, equivalently by the relations,  $\cos{(\theta_1/2)} = e^{-\frac{r}{2}}\sqrt{\sinh{r}}$ and $\phi_1 =\phi_2=-\varphi/2$, $\cos{(\theta_2/2)}=-e^{-\frac{r}{2}}\sqrt{\sinh{r}}$ in \eqref{Majorana}. Indeed, for $r = 0$, $ \ket{N}_c \ket{N}_d = \ket{2N}_b$, and $\Lim{N \rightarrow \infty} \ket{2N}_b \to \ket{\alpha=0}$ \cite{CommentSqueezing1, CommentSqueezing2}. Notice that a different choice of values for $\theta_1, \theta_2, \varphi_1, \varphi_2$ would lead to different squeezed states, as for instance, displaced squeezed states. The latter could be obtained by applying the $N \to \infty$ limit to states: $\ket{r,\varphi}_{\alpha}=\hat R((\theta,\phi)(\alpha,N) )\hat R((\eta,\chi)(r,\varphi))\ket{N}_{\theta_1,\phi_1}\ket{N}_b$.

The state $\ket{N}_{\theta_1,\phi_1}\ket{N}_b $ is a QNG and a non-classical SSRC state \cite{descamps2024superselectionrules}. It can be created from an initial classical state $\ket{2N}_b$ by the application of SNG operations, that we represent as $\hat U_{\cal S}(\theta_1, \theta_2=0, \phi_1,\phi_2=0)$ for the reference mode basis choice of \eqref{Majorana}. SNG are necessary for the creation of this state because, as discussed previously, SG operations are rotations $\hat R(\theta, \phi)\ket{2N}_b = \ket{2N}_{\theta,\phi}$, where $\ket{N}_{\theta,\phi}$ is still a classical (particle separable) state. Nevertheless, in the CV limit, $\hat U_{\cal S}(\theta_1, \theta_2=0, \phi_1,\phi_2=0)$ becomes a QG operation, since in this limit $\ket{2N}_b$ and $\ket{N}_{\theta_1,\phi_1}\ket{N}_b $ are QG states. This is an interesting remark that explicits how the physical consistency of the SNG and SG operations is lost in the CV limit, where a SNG operation becomes a QG one in \eqref{uni}. 

In addition, state $\ket{N}_{\theta_1,\phi_1}\ket{N}_b$ is a non-classical state presenting particle entanglement and a QNG state in the quadrature representation.  Nevertheless, $\ket{N}_{\theta_1,\phi_1}\ket{N}_b$ becomes a QG state in the CV limit, and with a proper choice of mode basis, showing once again that such a limit may hide intrinsic important properties of bosonic quantum states.

A key question we now address is how the measurement of SSRC observables can lead, in the appropriate limit, to results associated with CV {\it observables}, as the quadrature operators. In order to do so, we study as an example the SSRC uncertainty relation $\Delta \hat J_x \Delta \hat J_y \geq \frac{1}{2}\langle \hat J_z \rangle$. Using \eqref{Psi} in the CV limit, $\langle \hat J_z \rangle \approx N/2$ and  $\Delta \hat J_{x(y)} \approx \sqrt{N}\Delta \hat x_{0(\frac{\pi}{2})}$, where $\hat x_{\theta}=\frac{1}{\sqrt{2}}(e^{i\theta}\hat a + e^{-i\theta}\hat a^{\dagger})$, so the SSRC uncertainty relation becomes numerically equivalent to the Heisenberg uncertainty relation (details in Appendix \ref{App5}). Similar limits were obtained in different contexts, as in atomic systems \cite{RevModPhys.90.035005,PhysRevA.68.033821}, using operator transformations \cite{HP, HP2}.

Using the results developed in this section, we can now revisit a fundamental aspect of bosonic quantum states: the interpretation of errors, imperfections, and non-physical states in the CV limit using the SSRC representation. In the quadrature CV representation, physical squeezed states are defined as those that saturate the Heisenberg uncertainty principle and possess finite widths in both quadratures. These states have finite energy and are therefore physical. However, in the context of quantum information encoding - particularly for error correction schemes such as GKP codes - they are often compared to infinitely squeezed states - quadrature eigenstates -, which would have infinite precision in one quadrature and are thus unphysical due to their infinite energy. 

One conventional interpretation is to treat the finite width of physical squeezed states as a form of imperfection - viewing them as idealized, infinitely squeezed states corrupted by Gaussian displacement noise. From this perspective, that sets quadrature eigenstates as a privileged basis for being well adapted to the gates \eqref{uni}, any CV encoding is intrinsically noisy simply because the states are physical. Consequently, perfect information encoding appears to require non-physical states, and deviations from such idealized, unphysical targets are considered as errors. However, the proposed SSRC framework offers a completely different and physically grounded interpretation. In this formalism, the state $\ket{N}_{\theta_1,\phi_1}\ket{N}_b$, which corresponds to a squeezed state, is a perfectly well-defined physical {\it and logical} state. It can be consistently interpreted as qubit state in a particular encoding for any value of $\theta_1$ and $\phi_1$, as will be explicitly shown in the next section. From this perspective, it is no more or less ``imperfect" for information encoding, than the reference state $\ket{N}_a\ket{N}_b$ - an infinitely squeezed state -, which corresponds to the special case $\theta_1 = 0$, since there is a systematic way of constructing logical gates corresponding to the encoding of information for arbitrary values of $\theta_i, \phi_i$. As a matter of fact, $\ket{N}_{\theta_1,\phi_1}\ket{N}_b= \hat K\ket{N}_a\ket{N}_b$, where $\hat K$ is a SNG operation (again, it is easy to verify that no rotation, SG operation, can transform $\\ket{N}_{\theta_1,\phi_1}\ket{N}_b$ into $\ket{N}_a\ket{N}_b$), and any gate defined for the ``ideal" state $ket{N}_a\ket{N}_b$ can be readily adapted to an arbitrary encoding. Nonetheless, in the CV limit, the state $\ket{N}_a\ket{N}_b$ poses a problem: all terms in Eq.\eqref{squeezie} become equal, and the transformation into the canonical CV squeezed form of Eq.\eqref{nice} is no longer possible. This highlights a key limitation of the CV framework: certain perfectly physical states, such as $\ket{N}_a\ket{N}_b$, appear as non-physical or ill-defined. In contrast, the SSRC representation maintains a clear and physically consistent interpretation of all such states. Rather than viewing finite-width squeezed states as noisy approximations of non-physical ideals, they are treated as fundamentally valid elements of the physical and logical Hilbert space, capable of encoding information without invoking unphysical limits, simply by using the adapted associated logical gates. This reframing not only clarifies the role of imperfections and approximations as a change of encoding basis - a point that will be further developed in the next section -, but also avoids the paradoxical notion that physical realizability is itself a source of error.

\section{General formulation of information encoding}\label{SectionGeneral}

\subsection{A note on the measurement basis}

We begin with a note on measurement-related resources in bosonic quantum information. Measurements can themselves be considered quantum resources, as they are represented by positive operator-valued measures (POVMs). For example, homodyne and heterodyne detections are often treated as QG operations while photon-number resolved detections as QNG ones. However, this classification is an approximation that holds only in the weak coupling limit between light and matter. In general, photodetection is not inherently QG, and treating it as such neglects important physical subtleties, as in the previously discussed examples.

This point has been addressed as well in previous works \cite{Tyc_2004, PhysRevA.55.3195}, which describe homodyne detection as a QNG operation associated with POVM elements of the form $\hat{\Pi}_n = \ketbra{n}$. Moreover, as shown in the previous section, the simultaneous detection in different (orthogonal) modes constitutes a SNG operation. This is particularly relevant for protocols like BosonSampling \cite{v009a004}, where mode-resolved detection reveals the computational resources, or entanglement \cite{PlenioExtract, PhysRevX.10.041012} that may not be evident from the state alone \cite{descamps2024superselectionrules}.

From a computational perspective, the measurement basis can serve to define the computational basis itself. Accordingly, in this work, we assume that detection is performed in the computational basis. This means that measurement does not introduce additional computational or physical resources beyond those associated with the encoding — it reflects the same physical resources as the encoded states.

Therefore, our focus is on the requirements in terms of SG or SNG resources to define universal gates irrespectively of the encoding, rather than on states, which merely reflect an encoding choice. In this sense, the choice of encoding - and hence the state - is akin to defining a classicality criterion or selecting a pointer basis \cite{PhysRevD.24.1516}.

\subsection{Information encoding} 

We now provide general principles for quantum information encoding in bosonic fields and the required physical resources. Let us begin with the SSRC representation  \eqref{Psi}, \eqref{Majorana} of a single-mode CV state. We will use again that SG transformations - which become displacements (a QG operation) in the CV limit - leave the parameters ${\theta_k, \phi_k}$ in \eqref{Majorana} unchanged, up to a global mode basis change, $\hat a \to \hat R(\theta,\phi)\hat a\hat R^{\dagger}(\theta,\phi)$, $\hat b \to \hat R(\theta,\phi)\hat b\hat R^{\dagger}(\theta,\phi)$. In addition, SG operations do not change the intrinsic number of modes in \eqref{Psi}, \eqref{Majorana} \cite{TrepsModes}. We then conclude that SNG operations (generators at least quadratic in $\hat J_{\vec{n}}$) are necessary for increasing the number of modes and, consequently, for creating mode entanglement that cannot be separated by linear mode transformations \cite{PhysRevA.100.062129, PhysRevLett.130.090602, MattiaX} ({\it i.e.}, SNG operations create particle entanglement in first quantization). 

In the CCV limit, mode transformations correspond to displacements, and they can produce many (more than two) quasi-orthogonal states from an initial state thanks to phase locking. This could give the impression that an undetermined number of orthogonal states can be encoded in a single mode of the field in the CV limit. However, the SSRC representation sets a bound on this, given by the Hilbert space dimension $N+1$ \cite{Chryssomalakos_2018}. In addition, since the CV limit $N \to \infty$ is in fact  {\it relative} to physical constraints ($N \gg n_{\rm max}$), we can directly transpose theorems obtained in finite dimension, as the Solovay-Kitaev theorem \cite{dawson2005solovaykitaevalgorithm, PhysRevLett.126.190504}, to any bosonic encoding of arbitrary dimension, including in the CV limit.

Orthogonal modes can be used to define qubits (or qudits), and we now show how to create a bosonic quantum computer with an arbitrary encoding from a SSRC classical state $\ket{ N K}$, $K \in \mathbb{N}$ (hence, a Fock state with $NK$ photons). For this, we use a unitary SNG operation that increases the computational space, the relative phase unitary operator $\hat P$ \cite{PhysicsPhysiqueFizika.1.49, PhysRevA.48.4702,PhysRevA.63.063801},  $\hat P\ket{n}_a\ket{N-n}_b = \ket{n+1}_a\ket{N-n-1}_b$.  By applying $\hat P\ket{ N K}$, over different pairs of modes, the state $\ket{\cal I}=\prod_{k=1}^K \ket{0}_{a_k}\ket{N}_{b_k}$ can be produced, where $a_k, b_k$ are orthogonal modes. State $\ket{\cal I}$ can be used to define a $K$ qubit bosonic quantum computer where the $k$-th qubit is encoded using two orthogonal states in the form \eqref{Psi} in modes $a_k, b_k$. 
We now introduce a general SSRC model that encompasses all qubit bosonic encodings in modes $a_k, b_k$ (or one mode in the CV representation). Our model covers both the general $N$ case and the CV limit, and it can be directly extended to any possible encoding involving more modes.

A general encoding is expressed as $\ket{\overline 0}_{k}=\hat U^{(a_k,b_k)}_0\ket{N}_{b_k}$, $\ket{\overline 1}_k=\hat U^{(a_k,b_k)}_1\ket{N}_{a_k}$, $ {}_k\langle \overline 1 | \overline 0 \rangle_{k} = 0$  where $\hat U^{(a_k,b_k)}_{0(1)}=e^{i {\cal H}^{(k)}_{0(1)}(\hat J_{\vec{n}})}$, and ${\cal H}^{(k)}_{0(1)}(\hat J_{\vec{n}})$ is a polynomial of at most $N$-th degree in $\hat J_{\vec{n}}$, potentially involving different unit vectors $\vec{n}$  \cite{CommentEncoding}. For instance, for $\hat U^{(a_k,b_k)}_0=\hat U^{(a_k,b_k)}_1={\mathbb 1}$, we have an encoding in Fock states. It can be compared to the coherent state encoding in the CV limit ($|\alpha|^2 \ll N$), where $\ket{\overline 0} =\ket{N}_{-\alpha,N} \to \ket{-\alpha}$, $\ket{\overline 1}=\ket{N}_{\alpha,N} \to \ket{\alpha}$, {\it i.e.}, $\hat U^{(a_k,b_k)}_0=\hat R((\theta,\phi)(-\alpha,N) )$, $\hat U^{(a_k,b_k)}_1=\hat R((\theta,\phi)(\alpha,N) )\hat R(\frac{\pi}{2},0)$. This type of encoding is employed in superconducting-circuit-based quantum computers \cite{10.21468/SciPostPhysLectNotes.72, grimm:hal-02379475, guillaud_repetition_2019, MazyarParadigm, PhysRevLett.124.180502}.

Within the proposed encodings, single qubit gates, as the Hadamard gate $\hat H_k\ket{\overline 0}_k \rightarrow \ket{\overline +}_k=\frac{1}{\sqrt{2}}(\ket{\overline 0}_k+\ket{\overline 1}_k)$, $\hat H_k\ket{\overline 1}_k \rightarrow \ket{\overline -}_k=\frac{1}{\sqrt{2}}(\ket{\overline 0}_k-\ket{\overline 1}_k)$, require the application of SNG transformations \cite{PhysRevA.87.042315, PhysRevA.108.022424}. The Hadamard gate can produce Schrödinger cat states in the CV limit, as well as NOON states, for which $\theta_k \neq \theta_{k'}$, $\phi_k \neq \phi_{k'}$ for $k \neq k'$ in \eqref{Majorana} \cite{PhysRevA.87.053821, PhysRevA.70.023812}. However, in other encodings, such as GKP states \cite{gottesman_encoding_2001}, (where ${\cal H}^{(k)}_{0(1)}$ is at least quadratic in $\hat J_{\vec{n}}$), some single-qubit gates may correspond to SG operations (QG in the CV limit). Nevertheless, as we demonstrate in the Appendix \ref{App6}, with the exception of single photon encoding ($N=1$), for which linear mode manipulations are also manipulations of the encoding states, {\it all bosonic qubit encodings must include SNG gates in the set of local gates}.

Regardless of the choice of bosonic computational basis ({\it i.e.}, logical qubit encoding), restricting operations to local qubit gates is insufficient to achieve universality. Qubit entanglement is also necessary, and we demonstrate in Appendix \ref{App6} that within bosonic encodings, it can only be realized via SNG gates - these are also QNG entangling operations in the CV limit \cite{MattiaX}. A notable property of bosonic encodings, compared to abstract quantum computers, is that qubit entanglement necessarily requires SNG gates, which are often compared to non-Clifford gates, for consisting of operations introducing ``magic" to otherwise efficiently classically simulable bosonic architectures. 

Depending on the specific encoding and the available physical resources, a mode basis transformation of a local SNG gate may effectively convert it into a two-qubit SNG gate. Nevertheless, except in the special case of single-photon encoding, SNG gates are indispensable for implementing both local and non-local operations in bosonic quantum computing architectures.

In the context of CV quantum computing with optical systems, it is often claimed that beam splitter operations (or more generally, passive linear optics such as the SG gate) generate entanglement. While such operations can indeed produce mode entanglement, this corresponds merely to a change of basis in the mode space and does not generate qubit entanglement. Intuitively, this can be understood by noting that these operations are local in the first-quantization picture; they do not allow for conditional interactions between photons in different modes.

\section{Discussion and Conclusion}\label{SectionDiscussion}

We have introduced a unifying framework based on the SSRC representation that enables a clear and physically grounded identification of the necessary and sufficient resources for universal quantum computing with bosonic systems, irrespective of the encoding. Our approach is general, since it's valid across all bosonic regimes - from CV systems to fixed-photon-number encodings - without requiring a change of mathematical representation, any mapping, or additional physical interpretation.

Importantly, the SSRC framework works within a finite-dimensional Hilbert space of fixed total particle number, which, while arbitrarily large, remains bounded and conserved. This perspective eliminates several conceptual and technical limitations associated with the conventional CV formalism. Notably, it clarifies the origin of nonphysical states, the apparent disconnection between CV and discrete-variable bosonic models, and ambiguities in the interpretation of modes versus states in quantum optics. These issues, we argue, stem from treating a relative (CV) regime and a restricted subspace as if they were absolute and complete, respectively.

By grounding our analysis in a finite-dimensional and physically consistent representation, we have derived universal conditions for quantum computing that are independent of the choice of encoding - except in the case of single photon encoding. In contrast to prior approaches that rely on mapping physical systems to abstract mathematical constructs to reveal their computational structure, our method directly identifies the fundamental resources in the physical space of the system and on arbitrary states.

In addition, we identify SNG operations as necessary for qubit entanglement in bosonic systems. While SG operations merely induce mode transformations without generating particle interactions or increasing the informational complexity of the state, SNG operations correspond to particle interactions and modifies  the number of parameters required to describe a quantum state. We have shown in detail how many important features of the CV representation naturally emerge from the more general SSRC framework, stressing their dependency on the limit considered and how they lead to the disappearance of some fundamental properties of operations and states. This reveals that commonly used benchmarks for identifying operations and states as resources - even if we do not restrict to the quadrature basis but keep the CV limit - miss fundamental aspects of states and operations. As a result, simply changing the phase-space representation or computational basis while keeping the CV limit may be insufficient to consistently identify physically meaningful resources in bosonic systems and quantum protocols. Interesting perspectives that will be addressed in a future work include analyzing such results in terms of phase space representations for symmetric systems \cite{PhysRevA.49.4101, WignerNeg, PhysRevResearch.3.033134} or in different SSRC basis, as studied in \cite{AVourdas_2004}.

The SSRC representation may also serve as a valuable pedagogical tool by revealing the dependence of coherent states on the choice of a phase reference mode. From this perspective, the apparent continuity of bosonic systems does not stem from the states themselves—which form a discrete and finite basis—but rather from the structure of the modes. This reinterpretation, which contrasts with the conventional view, holds significant potential for enriching both undergraduate and graduate-level quantum physics education.

Beyond its conceptual advantages, the SSRC formalism bridges the gap between the photon-number-resolved and CV regimes in quantum optics and, more generally, bosonic systems. Its generality allows it to produce results that are valid across the entire bosonic spectrum, with direct implications for any quantum optical platform used for information processing \cite{GKPpropagating1, Review2, Reglade, Xanadu}. Additionally, the SSRC representation naturally extends to other physical systems composed of symmetric particles, such as spin ensembles and angular momentum systems \cite{RMPSym, PhysRevLett.127.010504, PhysRevA.108.022424}, further broadening its relevance.

Finally, this unified framework opens exciting perspectives for analyzing quantum error correction - which is beyond the scope of the present work - in both DV and CV regimes, including multimode encodings and rotational codes \cite{descamps:hal-04786822, PhysRevA.108.022424, PhysRevLett.127.010504, PRXQuantum.5.020355}. By enabling a systematic comparison and reinterpretation of quantum protocols across diverse platforms, the SSRC framework offers a unifying foundation with broad applicability to a wide class of physical systems.



\section*{Acknowledgements}

We acknowledge funding from the Plan France 2030 through the project ANR-22-PETQ-0006 and José Lorgeré, Chloé Lorgeré, Sara Ducci, Valentin Cambier, Luca Guidoni, Félicien Appas, Nicolas Fabre, Nicolas Treps and Isabelle Bouchoule for insightful discussions. This paper is dedicated to the memory of Prof. Marcelo Sá Corrêa, who first introduced P.M. to the binomial distribution and combinatorics in high school and sadly passed away during the preparation of this manuscript.

\appendix 

\section{Universality of SSRC operations}\label{App1}

In this section, we demonstrate that all states of the form
\begin{equation}
\label{eq:state}
\lvert \psi \rangle = \sum_{n=0}^{N} c_n \lvert n\rangle_a\lvert N-n\rangle_b
\end{equation}
can be generated by applying unitary operations built from polynomials of $\hat J_\pm$ to the initial state $\lvert 0\rangle_a\lvert N\rangle_b$. This construction is inspired by the universality result in~\cite{PhysRevLett.132.153601}, but here we adopt the two-mode Schwinger representation described in Section~I. Our approach naturally generalizes to the multi-mode electromagnetic field.

We aim to sequentially couple the states in Eq.~\eqref{eq:state} using unitary operators that are linear or quadratic (and eventually higher-order) functions of $\hat J_\pm$. We use the standard definitions:
\begin{align}
\hat J_+ &= \hat J_x + i\hat J_y = \hat a^\dagger \hat b, \nonumber \\
\hat J_- &= \hat J_x - i\hat J_y = \hat a \hat b^\dagger,\nonumber
\end{align}
which satisfy the actions
\begin{align}
\hat J_+ \lvert n\rangle_a\lvert N{-}n\rangle_b &= \sqrt{(n+1)(N{-}n)}\lvert n{+}1\rangle_a\lvert N{-}n{-}1\rangle_b, \nonumber \\
\hat J_- \lvert n\rangle_a\lvert N{-}n\rangle_b &= \sqrt{n(N{-}n{+}1)}\lvert n{-}1\rangle_a\lvert N{-}n{+}1\rangle_b,\nonumber
\end{align}
and the commutation relations
\begin{align}
[\hat J_x, \hat J_y] &= i\hat J_z, \quad\nonumber
[\hat J_+, \hat J_-] = 2\hat J_z = \hat b^\dagger \hat b - \hat a^\dagger \hat a.\nonumber
\end{align}

We begin with a first-order operation applied $M_1$ times:
\begin{align}
\left(e^{\alpha_1 \hat J_{+} - \alpha_1^{\ast} \hat J_{-}}\right)^{M_1} \lvert 0\rangle_a\lvert N\rangle_b.\nonumber
\end{align}
Assuming $|\alpha_1|^2 \ll 1$, we use the approximation
\begin{align}
e^{\alpha_1 \hat J_{+} - \alpha_1^{\ast} \hat J_{-}} \approx e^{\alpha_1 \hat J_{+}} e^{- \alpha_1^{\ast} \hat J_{-}},\nonumber
\end{align}
to obtain
\begin{align}
\left(e^{\alpha_1 \hat J_{+} - \alpha_1^{\ast} \hat J_{-}}\right)^{M_1} \lvert 0\rangle_a\lvert N\rangle_b \nonumber
&\approx \sum_{m=0}^{\infty} \frac{(M_1 \alpha_1 \hat J_+)^m}{m!} \lvert 0\rangle_a\lvert N\rangle_b \nonumber \\
&\approx \lvert \psi_0 \rangle + \frac{c_1}{c_0} \lvert \psi_1 \rangle,
\end{align}
where $\lvert \psi_n \rangle = \lvert n \rangle_a \lvert N-n \rangle_b$, and $M_1$ is chosen so that $M_1\alpha_1\sqrt{N} = c_1/c_0$.

Next, we apply a second-order operation $M_2$ times:
\begin{align}
\left(e^{\alpha_2 \hat J_{+}^2 - \alpha_2^{\ast} \hat J_{-}^2}\right)^{M_2} \left( \lvert \psi_0 \rangle + \frac{c_1}{c_0} \lvert \psi_1 \rangle \right).\nonumber
\end{align}
Again assuming $|\alpha_2|^2 \ll 1$, we approximate
\begin{align}
&\left(e^{\alpha_2 \hat J_{+}^2 - \alpha_2^{\ast} \hat J_{-}^2}\right)^{M_2} \nonumber
\approx e^{M_2 \alpha_2 \hat J_{+}^2} e^{-M_2 \alpha_2^{\ast} \hat J_{-}^2}, \nonumber\\
&\Rightarrow \lvert \psi \rangle 
\approx \lvert \psi_0 \rangle + \frac{c_1}{c_0} \lvert \psi_1 \rangle + \frac{c_2}{c_0} \lvert \psi_2 \rangle,\nonumber
\end{align}
with $M_2 \alpha_2 \sqrt{2N(N-1)} = c_2 / c_0$. 

Continuing this process recursively with operators of increasing order, we arrive at the full target state:
\begin{align}
\label{eq:universal_state}
&\left(e^{\alpha_N \hat J_{+}^N - \alpha_N^{\ast} \hat J_{-}^N}\right)^{M_N} \!\cdots\!
\left(e^{\alpha_1 \hat J_{+} - \alpha_1^{\ast} \hat J_{-}}\right)^{M_1}
\lvert 0\rangle_a\lvert N\rangle_b\nonumber \\
&= \frac{1}{c_0} \sum_{n=0}^N c_n \lvert \psi_n \rangle.
\end{align}

In \cite{PhysRevLett.132.153601} it is shown that the complexity of generating an arbitrary state scales with $N^4$. 
\subsection{Multimode universality}

The method outlined above can be extended to the multimode case, where $N$ photons are distributed across $K$ modes. The general symmetric state reads
\begin{align}
\label{eq:state_multimode}
\lvert\Psi\rangle &= \sum_{\vec{n}} c_{\vec{n}} \lvert \vec{n} \rangle 
= \hspace{-0.45cm} \sum_{\{n_i\}: \sum_i n_i = N} \hspace{-0.45cm} c_{n_1 \cdots n_K} \lvert n_1 \rangle_1 \cdots \lvert n_K \rangle_K,
\end{align}
where $\vec{n} = (n_1, n_2, \ldots, n_K)$ denotes the photon number configuration across modes, and $\sum_{i=1}^K n_i = N$. The coefficients $c_{\vec{n}}$ are the complex amplitudes corresponding to each Fock basis element.

To generate any state of the form~\eqref{eq:state_multimode}, we define angular momentum–like operators acting between pairs of modes:
\begin{subequations}
\label{eq:operators}
\begin{align}
\hat{J}_{x}^{(i,j)} &= \frac{1}{2}(\hat{a}_i^\dagger \hat{a}_j + \hat{a}_j^\dagger \hat{a}_i), \\
\hat{J}_{y}^{(i,j)} &= \frac{i}{2}(\hat{a}_i^\dagger \hat{a}_j - \hat{a}_j^\dagger \hat{a}_i), \\
\hat{J}_{z}^{(i,j)} &= \frac{1}{2}(\hat{a}_j^\dagger \hat{a}_j - \hat{a}_i^\dagger \hat{a}_i), \\
\hat{J}_{\pm}^{(i,j)} &= \hat{J}_x^{(i,j)} \pm i \hat{J}_y^{(i,j)},
\end{align}
\end{subequations}
with $1 \leq j < i \leq K$. We begin with the initial state $\lvert \Psi_0 \rangle = \lvert 0 \rangle_1 \cdots \lvert N \rangle_K$, where all photons occupy mode $K$.

Applying small-angle unitaries constructed from these operators:
\begin{align}
\prod_{j=1}^{K-1} \left(e^{\beta_j \hat{J}_+^{(K,j)} - \beta_j^* \hat{J}_-^{(K,j)}}\right)^{M_j} \lvert \Psi_0 \rangle,\nonumber
\end{align}
and assuming $|\beta_j|^2 \ll 1$, we expand the result:
\begin{align}
\lvert \Psi \rangle &\approx \lvert \Psi_0 \rangle 
+ \sqrt{N} \sum_{j=1}^{K-1} \beta_j M_j \lvert 0 \rangle_1 \cdots \lvert 1 \rangle_j \cdots \lvert N{-}1 \rangle_K.\nonumber
\end{align}
Choosing $M_j$ such that $\beta_j M_j \sqrt{N} = c_{(0,\cdots,1_j,\cdots,N{-}1)} / c_{(0,\cdots,N)}$ allows us to match the desired amplitudes. 

We then apply second-order and bilinear operations to generate higher-order photon redistributions:
\begin{align*}
&\prod_{j=1}^{K-1} \prod_{m>j}^{K-1}
\left(e^{\gamma_{j,m} \hat{J}_+^{(K,j)} \hat{J}_+^{(K,m)} - \gamma_{j,m}^* \hat{J}_-^{(K,j)} \hat{J}_-^{(K,m)}}\right)^{L_{j,m}}\nonumber \\
&\times \prod_{j=1}^{K-1} \left(e^{\delta_j (\hat{J}_+^{(K,j)})^2 - \delta_j^* (\hat{J}_-^{(K,j)})^2}\right)^{P_j} \lvert \Psi \rangle,\nonumber
\end{align*}
where $|\gamma_{j,m}|^2, |\delta_j|^2 \ll 1$. This produces terms such as:
\begin{align}
&\sum_{j=1}^{K-1} \frac{c_{(0,\cdots,2_j,\cdots,N{-}2)}}{c_{(0,\cdots,N)}} 
\lvert 0 \rangle_1 \cdots \lvert 2 \rangle_j \cdots \lvert N{-}2 \rangle_K \nonumber \\
+&\sum_{j < m}^{K-1} \frac{c_{(0,\cdots,1_j,1_m,\cdots,N{-}2)}}{c_{(0,\cdots,N)}} 
\lvert 0 \rangle_1 \cdots \lvert 1 \rangle_j \cdots \lvert 1 \rangle_m \cdots \lvert N{-}2 \rangle_K,\nonumber
\end{align}
with the amplitude conditions:
\begin{align}
P_j \delta_j \sqrt{2N(N{-}1)} &= \frac{c_{(0,\cdots,2_j,\cdots,N{-}2)}}{c_{(0,\cdots,N)}},\nonumber \\
L_{j,m} \gamma_{j,m} \sqrt{N(N{-}1)} &= \frac{c_{(0,\cdots,1_j,1_m,\cdots,N{-}2)}}{c_{(0,\cdots,N)}}.\nonumber
\end{align}

By recursively applying such unitary operations (composed of polynomials in $\hat{J}_\pm^{(i,j)}$), one can generate arbitrary symmetric multimode states of the form~\eqref{eq:state_multimode}.

\section{From the SSRC representation to coherent states}\label{App2}

In this section, we consider the Fock state $\ket{N}$ to extract a CV coherent state structure in the limit $N \to \infty$. 

We introduce two orthogonal modes with annihilation operators $\hat{a}$ and $\hat{b}$ such that $[\hat{a}, \hat{b}^\dagger] = 0$. Fixing $\alpha \in \mathbb{C}$, we define an $N$- and $\alpha$-dependent linear transformation $\hat{U}$ acting on the modes as follows:
\begin{align*}
    \hat{U} \hat{a}^\dagger \hat{U}^\dagger &= \sqrt{1 - \frac{|\alpha|^2}{N}}\, \hat{a}^\dagger - \frac{\alpha^*}{\sqrt{N}}\, \hat{b}^\dagger, \\
    \hat{U} \hat{b}^\dagger \hat{U}^\dagger &= \frac{\alpha}{\sqrt{N}}\, \hat{a}^\dagger + \sqrt{1 - \frac{|\alpha|^2}{N}}\, \hat{b}^\dagger.
\end{align*}

This transformation corresponds to the rotation $\hat{R}(\theta, \phi)$ defined in the main text, with parameters 
\[
\theta(\alpha,N) = 2\arccos\left(\sqrt{1 - \frac{|\alpha|^2}{N}}\right), \quad \phi(\alpha,N) = \arg(\alpha).
\]

A direct computation shows that 
\[
[\hat{U} \hat{a}^\dagger \hat{U}^\dagger, \hat{U} \hat{b}^\dagger \hat{U}^\dagger] = 0,
\]
confirming that $\hat{U}$ defines a valid unitary transformation. We now consider the state $\hat{U} \ket{N}_b$ in the limit $N \to \infty$:

\begin{subequations}
\begin{align}
    \hat{U} \ket{N}_b &= \frac{(\hat{U} \hat{b}^\dagger \hat{U}^\dagger)^N}{\sqrt{N!}} \ket{\emptyset}\nonumber  \\
    &= \frac{1}{\sqrt{N!}} \left( \frac{\alpha}{\sqrt{N}} \hat{a}^\dagger + \sqrt{1 - \frac{|\alpha|^2}{N}} \hat{b}^\dagger \right)^N \ket{\emptyset} \nonumber \\
    &= \frac{1}{\sqrt{N!}} \sum_{k=0}^N \binom{N}{k} \left( \frac{\alpha}{\sqrt{N}} \right)^k\nonumber  \\ 
    &\left( \sqrt{1 - \frac{|\alpha|^2}{N}} \right)^{N-k} (\hat{a}^\dagger)^k (\hat{b}^\dagger)^{N-k} \ket{\emptyset}\nonumber  \\
    &= \sum_{k=0}^N \binom{N}{k} \frac{\sqrt{k!} \sqrt{(N-k)!}}{\sqrt{N!}} \left( \frac{\alpha}{\sqrt{N}} \right)^k \nonumber \\
    &\left( \sqrt{1 - \frac{|\alpha|^2}{N}} \right)^{N-k} \ket{k}_a \ket{N-k}_b\nonumber  \\
    &= \sum_{k=0}^N \sqrt{ \frac{N(N-1)\cdots(N-k+1)}{k!} } \left( \frac{\alpha}{\sqrt{N}} \right)^k \nonumber  \\
    &\left( \sqrt{1 - \frac{|\alpha|^2}{N}} \right)^N \left( \sqrt{1 - \frac{|\alpha|^2}{N}} \right)^{-k} \ket{k}_a \ket{N-k}_b \nonumber \\
    &\simeq \sum_{k=0}^N \frac{1}{\sqrt{k!}} \alpha^k e^{-|\alpha|^2/2} \ket{k}_a \ket{N-k}_b \nonumber \\
    &= e^{-|\alpha|^2/2} \sum_{k=0}^N \frac{\alpha^k}{\sqrt{k!}} \ket{k}_a \ket{N-k}_b.
\end{align}
\end{subequations}

The approximation is taken term by term with fixed $k$ as $N \to \infty$, using:
\begin{align*}
    N(N-1)\cdots(N-k+1) &\simeq N^k, \\
    \left( \sqrt{1 - \frac{|\alpha|^2}{N}} \right)^N &\simeq e^{-|\alpha|^2/2}, \\
    \left( \sqrt{1 - \frac{|\alpha|^2}{N}} \right)^k &\simeq 1.
\end{align*}

\section{Displacement Operator in SSRC Representation}\label{App3}

In the previous section, the transformation $\hat U$ maps $\ket{N}_b$ (the SSRC equivalent of vacuum in CV $\ket{0}$) to a state which becomes the SSRC version of a coherent state of amplitude $\alpha$ as $N\to\infty$. The displacement operator $\hat D(\alpha) = e^{\alpha\hat a^\dagger - \alpha^*\hat a}$ maps the vacuum to the coherent state. We now compare the action of $\hat D(\alpha)$ to $\hat U$ on the state $\ket{k}_a\ket{N-k}_b$ in the limit of large $N$.

We use the Baker--Campbell--Hausdorff formula:
\begin{equation*}
    \hat D(\alpha) = e^{-\lvert\alpha\rvert^2/2}e^{\alpha\hat a^\dagger}e^{-\alpha^*\hat a}.\nonumber
\end{equation*}

With ladder operator actions:
\begin{align*}
    \hat a^\dagger\ket{l} = \sqrt{l+1}\ket{l+1}, &\quad \hat a\ket{l} = \sqrt{l}\ket{l-1},\nonumber \\
    (\hat a^\dagger)^j\ket{l} = \frac{\sqrt{(j+l)!}}{\sqrt{l!}}\ket{j+l}, &\quad (\hat a)^j\ket{l} = \frac{\sqrt{l!}}{\sqrt{(l-j)!}}\ket{l-j}.\nonumber
\end{align*}

Then:
\begin{subequations}
\begin{align}
    \hat D(\alpha)\ket{k} &= e^{-\lvert\alpha\rvert^2/2}\sum_{l=0}^k \sum_{j=0}^\infty \frac{(-\alpha^*)^{k-l}\alpha^j}{(k-l)!j!}\frac{\sqrt{k!}\sqrt{(l+j)!}}{l!}\ket{l+j}.
\end{align}
\end{subequations}

Now compute $\hat U\ket{k}_a\ket{N-k}_b$:
\begin{subequations}
\begin{align}
    \hat U\ket{k}_a\ket{N-k}_b &= \sum_{l=0}^k\sum_{j=0}^{N-k}\binom{k}{l}\binom{N-k}{j}\times \nonumber \\
    &\frac{\sqrt{(j+l)!}\sqrt{(N-j-l)!}}{\sqrt{k!}\sqrt{(N-k)!}}\left(1{-}\frac{\lvert\alpha\rvert^2}{N}\right)^{\frac{(N-j-l-k)}{2}} \nonumber \\
    &\frac{\alpha^j(-\alpha^*)^{k-l}}{N^{(j+k-l)/2}}\ket{j+l}_a\ket{N-j-l}_b.
\end{align}
\end{subequations}

Term-by-term, we compare with $\hat D(\alpha)\ket{k}$ for fixed $j,l$ as $N\to\infty$:
\begin{itemize}
    \item Power of $\alpha$ and $\alpha^*$ matches.
    \item $(1{-}\lvert\alpha\rvert^2/N)^{(N-j-l-k)/2} \to e^{-\lvert\alpha\rvert^2/2}$.
    \item Combinatorics: $\frac{\sqrt{(N-k)!}\sqrt{(N-j-l)!}}{(N-k-j)!N^{(j+k-l)/2}}\to 1$.
\end{itemize}

Thus, $\hat U\ket{k}_a\ket{N-k}_b \to \hat D(\alpha)\ket{k}$ in the SSRC representation as $N\to\infty$.

\section{The SSRC representation and squeezed states in the CV limit}\label{App4}

In the CV representation the squeezed vacuum state reads:
\begin{equation}
\begin{split}
\ket{r,\phi} &= \frac{1}{\sqrt{\cosh(r)}} \sum_{k=0}^\infty 
\Bigl(-e^{i\phi}\tanh(r)\Bigr)^k 
\frac{\sqrt{2k!}}{2^k k!} \ket{2k}_a \\
&= \frac{1}{\sqrt{\cosh(r)}} \sum_{k=0}^\infty 
\Bigl(-e^{i\phi}\tanh(r)\Bigr)^k \frac{1}{2^k k!} \\
&\quad \times (\hat a^\dagger)^{2k} \ket{\emptyset}.
\end{split}
\end{equation}

In the SSRC formulation,
\begin{align}
\ket{r,\phi} &\simeq \frac{1}{\sqrt{\cosh(r)}}
\sum_{k=0}^\infty 
\Bigl(-e^{i\phi}\tanh(r)\Bigr)^k
\frac{\sqrt{2k!}}{2^k k!} \ket{2k}_a \otimes \ket{2(N-k)}_b,\nonumber\\
&= \frac{1}{\sqrt{\cosh(r)}}
\sum_{k=0}^\infty 
\Bigl(-e^{i\phi}\tanh(r)\Bigr)^k \frac{1}{2^k k! \sqrt{(2(N-k))!}} \nonumber \\
&\quad \times (\hat a^\dagger)^{2k} (\hat b^\dagger)^{2(N-k)} \ket{\emptyset}.\nonumber
\end{align}

We write
\[
\ket{r,\phi} = \lim_{N\to\infty} \frac{1}{A} (\hat c^\dagger)^N (\hat d^\dagger)^N \ket{\emptyset},\nonumber
\]
where $A$ is a normalization factor that depends on $r,N$, and
\begin{align}
\hat c^\dagger &= e^{-r/2} \Bigl(
e^{i\phi/2}\sqrt{\sinh(r)} \hat a^\dagger + \sqrt{\cosh(r)} \hat b^\dagger
\Bigr),\nonumber \\
\hat d^\dagger &= e^{-r/2} \Bigl(
- e^{i\phi/2}\sqrt{\sinh(r)} \hat a^\dagger + \sqrt{\cosh(r)} \hat b^\dagger
\Bigr).\nonumber
\end{align}

The operators satisfy
\[
[\hat c, \hat c^\dagger] = [\hat d, \hat d^\dagger] = 1,
\quad s := [\hat c, \hat d^\dagger] = e^{-r}.
\]

Expanding,
\begin{multline}\label{devel2}
\frac{1}{A} (\hat c^\dagger)^N (\hat d^\dagger)^N \ket{\emptyset} = 
\frac{e^{-Nr}}{A} \Bigl[
e^{i\phi/2}\sqrt{\sinh(r)} \hat a^\dagger + \sqrt{\cosh(r)} \hat b^\dagger
\Bigr]^N \nonumber \\
\times \Bigl[
- e^{i\phi/2}\sqrt{\sinh(r)} \hat a^\dagger + \sqrt{\cosh(r)} \hat b^\dagger
\Bigr]^N \ket{\emptyset}.\nonumber
\end{multline}

Using binomial expansion,
\begin{multline}
= \frac{e^{-Nr}}{A} \sum_{k=0}^N \binom{N}{k} (- e^{i\phi} \sinh(r))^k 
(\cosh(r))^{N-k}\nonumber \\
\times (\hat a^\dagger)^{2k} (\hat b^\dagger)^{2(N-k)} \ket{\emptyset}.\nonumber
\end{multline}

Rewrite using $\tanh(r)$,
\begin{multline}
= \frac{(e^{-r} \cosh(r))^N}{A} \sum_{k=0}^N \binom{N}{k} (- e^{i\phi} \tanh(r))^k \nonumber\\
\times (\hat a^\dagger)^{2k} (\hat b^\dagger)^{2(N-k)} \ket{\emptyset}.\nonumber
\end{multline}

Expressing Fock states,
\begin{multline}
= \frac{(e^{-r} \cosh(r))^N}{A} \sum_{k=0}^N (- e^{i\phi} \tanh(r))^k \nonumber\\
\frac{N! \sqrt{(2k)! (2(N-k))!}}{k! (N-k)!} \nonumber\\
\times \ket{2k}_a \otimes \ket{2(N-k)}_b.\nonumber
\end{multline}

For the normalization factor,
\begin{multline}
A^2 = (e^{-r} \cosh(r))^{2N} \sum_{k=0}^N (\tanh(r))^{2k}\nonumber \\
\times \frac{(N!)^2 (2k)! (2(N-k))!}{(k! (N-k)!)^2}.\nonumber
\end{multline}

In the CV limit,
\begin{multline}
A \to \frac{(2 e^{-r} \cosh(r))^N N!}{(\pi N)^{1/4}} \sqrt{
\sum_{k=0}^{n_{\mathrm{max}}} (\tanh(r))^{2k} \frac{(2k)!}{2^{2k} (k!)^2}
} = \nonumber\\ 
\frac{(2 e^{-r} \cosh(r))^N N!}{(\pi N)^{1/4}} \sqrt{\cosh r}.\nonumber
\end{multline}

Hence the normalized state corresponds to the vacuum squeezed state.

\section{Quadrature operators from the SSRC formulation}\label{App5}

	
		
	
In this section, we show how the quadrature field operators emerge from the SSRC formulation.  
For this, consider a general 2-mode SSRC state 
\begin{equation}
\ket{\Psi} = \sum_{k=0}^N c_k \ket{k}_a \ket{N-k}_b,\nonumber
\end{equation}
where modes $a$ and $b$ are orthogonal, and the photon number $N$ can be taken sufficiently large such that in the CV-limit we obtain the state
\begin{equation}
\ket{\psi} = \sum_{k=0}^\infty c_k \ket{k}_a.\nonumber
\end{equation}
This connection between the SSRC representation and the CV-limit has been specified in the main text and denoted as
\[
\ket{\Psi} \rightarrow \ket{\psi}.\nonumber
\]

Let us consider the $N$-dependent operator 
\begin{equation}
\hat{A}(N) = \frac{1}{\sqrt{N}} \hat{a} \hat{b}^\dagger,\nonumber
\end{equation}
where $N$ is the number of photons.
When applied to the state $\ket{\Psi}$, we have	
\begin{equation}
\hat{A}(N) \ket{\Psi} = \sum_{k=0}^N c_k 
\sqrt{k \left( 1 - \frac{k-1}{N} \right)} 
\ket{k-1}_a \ket{N-k+1}_b.\nonumber
\end{equation}

Applying the CV limit procedure introduced in the main text, which consists in taking the limit $N \to +\infty$ in two steps, 
\begin{align}
& \lim_{n_{\mathrm{max}} \to +\infty} \nonumber \\
&\sum_{k=0}^{n_{\mathrm{max}}} 
\lim_{N \to \infty} \left[ c_k 
\sqrt{k \left( 1 - \frac{k-1}{N} \right)} \right] 
\ket{k-1}_a \ket{n_{\mathrm{max}} - k + 1}_b \nonumber \\
&= \lim_{n_{\mathrm{max}} \to +\infty} \sum_{k=0}^{n_{\mathrm{max}}} 
c_k \sqrt{k} \,
\ket{k-1}_a \ket{n_{\mathrm{max}} - k + 1}_b,\nonumber
\end{align}
and then disentangling mode $a$ and $b$, we finally obtain
\begin{equation}
\hat{A}(N) \ket{\Psi} \rightarrow \sum_{k=0}^\infty c_k \sqrt{k} \ket{k-1}_a 
= \hat{a} \ket{\psi},\nonumber
\end{equation}
where $\hat{a}$ is the annihilation operator of mode $a$.
Since this relation holds for every state $\ket{\Psi}$ which in the CV limit gives $\ket{\psi}$, 
we write the CV limit of the operator as
\begin{equation}
\hat{A}(N) \rightarrow \hat{a}.\nonumber
\end{equation}

Similarly,
\begin{equation}
\hat{A}^\dagger(N) = \frac{1}{\sqrt{N}} \hat{a}^\dagger \hat{b} \rightarrow \hat{a}^\dagger.\nonumber
\end{equation}

Using $\hat{A}(N)$ and $\hat{A}^\dagger(N)$, we define the operator
\begin{equation}
\hat{Q}(N, \phi) = \frac{1}{\sqrt{2}} \left[ e^{-i\phi} \hat{A}(N) + e^{i\phi} \hat{A}^\dagger(N) \right], 
\quad \phi \in [0,2\pi],\nonumber
\end{equation}
which in the CV limit gives the quadrature operator:
\begin{equation}
\hat{Q}(N,\phi) \rightarrow \hat{q}(\phi) 
= \frac{1}{\sqrt{2}} \left[ e^{-i\phi} \hat{a} + e^{i\phi} \hat{a}^\dagger \right].\nonumber
\end{equation}

In particular, for $\phi=0$ and $\phi = \frac{\pi}{2}$, we have
\begin{align}
\hat{Q}(N,0) &= \sqrt{\frac{2}{N}} \hat{J}_x \rightarrow \hat{x} = \frac{1}{\sqrt{2}} (\hat{a} + \hat{a}^\dagger), \\
\hat{Q}(N,\frac{\pi}{2}) &= -\sqrt{\frac{2}{N}} \hat{J}_y \rightarrow \hat{p} 
= \frac{i}{\sqrt{2}} (\hat{a}^\dagger - \hat{a}).\nonumber
\end{align}

Finally, the commutation relation $[\hat{x}, \hat{p}] = i$ is obtained from SSRC using the commutation relation $[\hat{J}_x, \hat{J}_y] = i \hat{J}_z$.
Indeed, starting with
\begin{equation}
[\hat{Q}(N,0), \hat{Q}(N,\frac{\pi}{2})] = -\frac{2i}{N} \hat{J}_z,\nonumber
\end{equation}
and noting that
\begin{align}
&-\frac{2}{N} \hat{J}_z \ket{\Psi} = \sum_{k=0}^N \left( 1 - \frac{2k}{N} \right) c_k \ket{k}_a \ket{N-k}_b 
 \nonumber \\
&\rightarrow\sum_{k=0}^\infty c_k \ket{k}_a = \ket{\psi},\nonumber
\end{align}
we can write the CV limit of the commutation relation as
\begin{equation}
[\hat{Q}(N,0), \hat{Q}(N,\frac{\pi}{2})] = -\frac{2i}{N} \hat{J}_z \rightarrow [\hat{x}, \hat{p}] = i.\nonumber
\end{equation}

\section{Details on the calculations on generic bosonic encodings and mode transformations}\label{App6}

We start, for simplicity, from the  encoding $\ket{\overline 0}=\ket{N}_b$, $\ket{\overline 1}=\hat U_1\ket{N}_a$, $ \langle \overline 1 | \overline 0 \rangle = 0$. 

Within this encoding, state $\ket{\overline 0}$ is a single mode state in mode $b$, and state $\ket{\overline 1}$ is a two-mode state involving modes $a$ and $b$. Notice that, in this description, mode $a$ or $b$ can be a phase reference, and we can obtain a single mode CV encoding by using the methods presented on the main text, so we will present here a general description for arbitrary $N$. 


We start by showing that we cannot implement the (qubit) local gates of the universal set by using only Gaussian SSRC gates within this encoding, {\it i.e.}, local (in the encoding) SNG gates are also necessary. We then generalize our results and show that the same holds for any encoding, except for the single photon one. In this section we will always consider the general SSRC representation. 

To show our main result, we analyze the local gates of the universal set formed by {\it qubit} rotations (which can be different from SSRC rotations), that we define as ${\cal R}_x(\theta_x),{\cal R}_y(\theta_y),{\cal R}_z(\theta_z)$, where ${\cal R}_{\alpha}(\theta_{\alpha})$ is a rotation of an angle $\theta_{\alpha}$ around the axis $\alpha$ of the Bloch sphere (that again, may or may not correspond to the axis of rotation of $\hat R(\theta,\phi)$ as defined in the main text). 

We start by supposing that it is possible to implement the qubit rotation ${\cal R}_y(\theta_y)$ using a SG transformation $\hat R(\theta,\phi)$ in an encoding where $\ket{\overline 0}=\ket{N}_b$, where we are using eigenstates of the effective Pauli $\hat Z = \frac{1}{2}\left ( \ket{\overline 1}\bra{\overline 1}-\ket{\overline 0}\bra{\overline 0}\right )$ operator for the logical qubits as the computational basis. We should have then have
\begin{eqnarray}\label{rot}
&&\hat R(\theta,\phi)\ket{N}_b=(\cos{\frac{\theta}{2}})^N\ket{N}_b+\nonumber \\
&&\sum_{n=1}^N\binom{N}{n}^{\frac{1}{2}}( \cos{\frac{\theta}{2}})^{N-n}(e^{i\phi}\sin{\frac{\theta}{2}})^n\ket{n}_a\ket{N-n}_b=\nonumber \\
&&\cos{\left (\frac{\theta_y}{2} \right)}\ket{\overline 0}+\sin{\left (\frac{\theta_y}{2} \right)}\ket{\overline 1}={\cal R}_y(\theta_y)\ket{\overline 0}, \nonumber
\end{eqnarray}

For \eqref{rot} to be true, we have that $(\cos{\frac{\theta}{2}})^N=\cos{\left (\frac{\theta_y}{2} \right)}$. In addition, if $\hat R(\theta,\phi)$ is a qubit rotation, 
\begin{eqnarray}\label{um}
&&\ket{\overline 1}=\left (\sin{\left (\frac{\theta_y}{2} \right ) }\right )^{-1}\nonumber \\
&&\sum_{n=1}^N\binom{N}{n}^{\frac{1}{2}}( \cos{\frac{\theta}{2}})^{N-n}(e^{i\phi}\sin{\frac{\theta}{2}})^n\ket{n}_a\ket{N-n}_b=\nonumber \\
&&\left (\sin{\left (\frac{\theta_y}{2} \right ) }\right )^{-1}\hat R(\theta,\phi)\ket{N}_b-\left (\tan{\left (\frac{\theta_y}{2} \right ) }\right )^{-1}\ket{N}_b.\nonumber
\end{eqnarray} 
Hence, according to \eqref{um}, the definition of state $\ket{\overline 1}$ depends for $N>1$, on the rotation angle $\theta_y$, which is clearly not possible: for a fixed $N$ and for each $\theta_y$, we can in principle find a value of $\theta$  such that $\hat R(\theta,\phi)\ket{N}_b$ corresponds to a  rotation of an angle $\theta_y$ of the logical qubit. A convenient choice of $\theta$ is such that $\cos{\frac{\theta}{2}} = (\cos\frac{\theta_y}{2})^{1/N}$, with $\theta_y \in [0,\pi]$, meaning that Eq. \eqref{rot} is fulfilled and  $\ket{\overline 1}$ is given by the expression in the first line of Eq. \eqref{um}. Then we can choose $\theta_y = \pi$, that leads to $\cos{\frac{\theta}{2}} = (\cos\frac{\theta_y}{2})^{1/N} = 0$, and  $\theta = \theta_y = \pi$, so the resulting logical qubit should be equal to $\ket{\overline 1}$: $\hat R(\pi,0)\ket{N}_b = {\cal R}_y(\pi)\ket{\overline 0}=\ket{\overline 1}$ . Now, using Eq. \eqref{um}, we have that $\hat R(\pi,0) \ket{N}_b = \ket{N}_a$, which is clearly a contradiction, except for $N=1$. Indeed, for $N=1$, we can choose $\theta = \theta_y$ and define a rotation with a SG operation.

In order to further illustrate the impossibility of defining $\ket{\overline 1}$, we can notice that for $\hat R(\theta,\phi)$ to be a qubit rotation, we must also have 
\begin{eqnarray}\label{rot2}
&&\hat R(\theta,\phi)\ket{\overline 1}=\cos{\left (\frac{\theta_y}{2} \right)}\ket{\overline 1}-\nonumber \\
&&\sin{\left (\frac{\theta_y}{2} \right)}\ket{\overline 0}=\left (\sin{\left (\frac{\theta_y}{2} \right ) }\right )^{-1}\hat R^2(\theta,\phi)\ket{N}_b-\nonumber \\
&&\left (\tan{\left (\frac{\theta_y}{2} \right ) }\right )^{-1}\hat R(\theta,\phi)\ket{N}_b\nonumber \\
&&=-\cos{\left (\frac{\theta_y}{2} \right)}\ket{\overline 1}-\frac{\cos^2{\left (\frac{\theta_y}{2} \right)}}{\sin{\left (\frac{\theta_y}{2} \right)}}\ket{\overline 0}+\left (\sin{\left (\frac{\theta_y}{2} \right ) }\right )^{-1}\times \nonumber \\
&&\sum_{n=0}^N\binom{N}{n}^{\frac{1}{2}}((\cos{\frac{\theta}{2}})^2-e^{i\phi}(\sin{\frac{\theta}{2}})^2)^{N-n}(e^{i\phi}+e^{2i\phi})^n\times \nonumber \\
&&(\sin{\frac{\theta}{2}}\cos{\frac{\theta}{2}})^n\ket{n}_a\ket{N-n}_b).\nonumber
\end{eqnarray}
Again, $\hat R(\theta,\phi)$ cannot be a qubit rotation: by studying, for instance, the example of $\theta_y=\frac{\pi}{2}$, we see that in this case, we must have $(\cos{\frac{\theta}{2}})^N=\frac{1}{\sqrt{2}}$. At the same time, if $\theta_y=\pi/2$, \eqref{rot2} imposes that 
\begin{align} 
&\sum_{n=0}^N\binom{N}{n}^{\frac{1}{2}}((\cos{\frac{\theta}{2}})^2-\nonumber \\
&e^{i\phi}(\sin{\frac{\theta}{2}})^2)^{N-n}(e^{i\phi}+e^{2i\phi})^n(\sin{\frac{\theta}{2}}\cos{\frac{\theta}{2}})^n\ket{n}_a\ket{N-n}_b \nonumber \\
&=\ket{\overline 1},\nonumber
\end{align}
that can only be orthogonal to $\ket{\overline 0}$  if  $\theta=\pi/2$, which is clearly impossible, since it is in contradiction with $(\cos{\frac{\theta}{2}})^N=\frac{1}{\sqrt{2}}$, unless  $N=1$.  Since the qubit rotations of the particular angle of $\theta_y=\pi/2$ cannot be implemented with SG operations, we conclude that physical operations representing the qubit rotations ${\cal R}_y(\theta_y)$ cannot be implemented with SG operations, so the physical implementation of ${\cal R}_y(\theta_y)$ requires SNG gates. 
So, a partial conclusion is that if state $\ket{N}_b$ is part of the encoding space, one cannot define a $y$ qubit rotation gate with SG operations, so SNG operations are necessary.

 We now define as $\hat {\cal R}^{\nu}_{\alpha}[\hat U_0,\hat U_1](\theta_{\alpha})$  ($\alpha=x,y,z$) the SSRC physical operations implementing the local operations $\hat {\cal R}_{\alpha}(\theta_{\alpha})$ of the universal set of an abstract quantum computer on a given encoding. Each encoding is  univocally characterized by the unitary operators $\hat U_0, \hat U_1$ ($\ket{\overline 0}=\hat U_0\ket{N}_b$ and $\ket{\overline 1}=\hat U_1\ket{N}_a$), and  $\nu$ correspond to different SSRC operations that act as a same gate on the a given encoding. We have then that if $\hat {\cal R}^{\nu}_{\alpha}[\hat U_0,\hat U_1](\theta_{\alpha})$ is SG,  there exist some $\theta$, $\phi$ such that $\hat R(\theta,\phi) =  \hat {\cal R}^{\nu}_{\alpha}[\hat U_0,\hat U_1](\theta_{\alpha})$, and $\theta,\phi$ will depend on $\theta_{\alpha}$ and $\alpha$, but they will not be necessarily identical nor univocal. Using this notation, the partial conclusion above can then be stated as $\hat {\cal R}^{\nu}_{y}[\mathbb{1},\hat U_1](\theta_{\alpha}) \in {\cal N}$, where ${\cal N}$ is the set of SNG operations (and ${\cal G}$ the set of SG ones).   
 
 We can generalize our results to arbitrary encodings by asking ourselves whether there exist sets 
 \begin{align}
&S_{\nu}[ \hat U_0, \hat U_1]=\nonumber\\
&\{\hat {\cal R}^{\nu}_{x}[\hat U_0,\hat U_1](\theta_x),\hat {\cal R}^{\nu}_{y}[\hat U_0,\hat U_1](\theta_x),\hat {\cal R}^{\nu}_{z}[\hat U_0,\hat U_1](\theta_x)\}\nonumber
\end{align}
for a given encoding $(\hat U_0,\hat U_1)$ where all the SSRC operations $\hat {\cal R}^{\nu}_{\alpha}[\hat U_0,\hat U_1](\theta_{\alpha})$ representing qubit local operations $\hat {\cal R}^{\nu}_{\alpha}(\theta_{\alpha})$ are Gaussian. A qubit rotation around $y$ on an arbitrary encoding $\hat U_0, \hat U_1$ reads
\begin{align} 
&\hat {\cal R}^{\nu}_{y}[\hat U_0,\hat U_1](\theta_y)\hat U_0\ket{N}_b =\nonumber \\
&\cos{\left (\frac{\theta_y}{2}\right )}\hat U_0\ket{N}_b+\sin{\left (\frac{\theta_y}{2}\right )}\hat U_1\ket{N}_a,\nonumber
\end{align}
or 
\begin{align}
&\hat U^{\dagger}_0\hat {\cal R}^{\nu}_{y}[\hat U_0,\hat U_1](\theta_y)\hat U_0\ket{N}_b = \nonumber \\
&\cos{\left (\frac{\theta_y}{2}\right )}\ket{N}_b+\sin{\left (\frac{\theta_y}{2}\right )}\hat U'_1\ket{N}_a=\hat {\cal R}^{\nu'}_{y}[\mathbb{1},\hat U'_1](\theta_y)\ket{N}_b, \nonumber
\end{align}
so for every gate in the encoding $(\hat U_0,\hat U_1)$, there is a gate in the encoding $(\mathbb{1},\hat U'_1)$ and, in order to have  $\hat {\cal R}^{\nu}_{y}[\hat U_0,\hat U_1](\theta_y) \in {\cal G}$, we must have $\hat U_0 \in {\cal N}$ (since $\hat {\cal R}^{\nu'}_{y}[\mathbb{1},\hat U'_1](\theta_y)$ acts as a qubit rotation in the encoding $(\mathbb{1},\hat U_1)$, so it must be SNG). This result generalizes straightforwardly to the encodings  where $(\hat U_0 \in {\cal G}, \hat U_1 \in {\cal G}\cup {\cal N})$ and $(\hat U_0\in {\cal G}\cup {\cal N}, \hat U_1  \in {\cal G})$.

 Hence,  $\hat {\cal R}^{\nu}_{y}[\hat U_0,\hat U_1](\theta_y)$ can only be a SG operation if $\hat U_0$ and $\hat U_1$ are SNG. However, not all SNG operations would fit, since  $\hat {\cal R}^{\nu}_{y}[\hat U_0,\hat U_1](\theta) \in {\cal N}$ as well for encodings where $\ket{\overline 0}$ is one of $\hat J_z$ eigenstates (this can be shown in the same way as for the case $\hat U_0=\mathbb{1}$, {\it i.e.}, by computing the coefficients of an hypothetic $\ket{\overline 1}$ state), and also when $U_0$ and $U_1$ are diagonal on the $z$ basis (since in this case the operations act as a global phase).  We must then focus on encodings $\hat U_0\ket{N}_b$ with $\hat U_0 \in {\cal N}$ that are in the form of superpositions $\ket{\overline 0}=\sum_{n=0}^N c_n\ket{n}_a\ket{N-n}_b$ and ask whether it is possible for the SSRC operations representing the other qubit rotations, say,  $\hat {\cal R}^{\nu}_{z}[\hat U_0,\hat U_1](\theta_z) \in {\cal G}$ for the SNG encodings where $ \hat {\cal R}^{\nu}_{y}[\hat U_0,\hat U_1](\theta_y) \in {\cal G}$.

We can check this by verifying if it is possible to implement $\hat {\cal R}_{z}(\theta_z) \ket{\overline 0} =e^{i\theta_z}\ket{\overline 0}$ and $\hat {\cal R}_{z}(\theta_z) \ket{\overline 1} = e^{-i\theta_z}\ket{\overline 1}$ with a SG operation  $\hat R(\theta,\phi)$, {\it i.e.}, if $\hat {\cal R}^{\nu}_{z}[\hat U_0,\hat U_1](\theta_z)  \in {\cal G}$ or, equivalently, $\hat {\cal R}^{\nu}_{z}[\hat U_0,\hat U_1](\theta_z) = \hat R(\theta,\phi)$. 

For this, we should then have $\hat R(\theta,\phi)\hat U_0\ket{N}_b = e^{i\theta_z}\hat U_0\ket{N}_b$, that can be written as 
\begin{align}
&e^{i\eta \hat J_z}\hat R^{\dagger}(\theta',\phi')\hat U_o\hat R(\theta',\phi')\hat R^{\dagger}(\theta',\phi')\ket{N}_b=\nonumber \\
&e^{i\varphi}\hat R^{\dagger}(\theta',\phi')\hat U_0\hat R(\theta',\phi')\hat R^{\dagger}(\theta',\phi')\ket{N}_b,\nonumber
\end{align}
where we expressed $\hat R(\theta,\phi)=\hat R(\theta',\phi')e^{i\eta \hat J_z}\hat R^{\dagger}(\theta',\phi')$. This is a simple mode rotation that changes the mode basis $a \to c$, $b \to d$ and, in principle, the coefficients $c_n \to c'_n$ if we choose to express the rotated state in the mode basis $a$, $b$. However, the rotated SNG operation $\hat R^{\dagger}(\theta',\phi')\hat U_0\hat R(\theta',\phi')$ remains SNG. It means that we can simply check whether it is possible to have 
\begin{align}\label{one}
&e^{i\eta \hat J_z}\sum_{n=0}^N c'_n \ket{n}_a\ket{N-n}_b=\sum_{n=0}^N e^{i\eta (2n-N)}c'_n \ket{n}_a\ket{N-n}_b\nonumber \\
&=e^{i\theta_z}\sum_{n=0}^N c'_n \ket{n}_a\ket{N-n}_b,\nonumber
\end{align}
with $c'_n \neq 1~ \forall ~n$. We call the first $c'_{n} \neq 0$ $c'_{n_0}$, and the condition for $e^{i\eta \hat J_z}$ to correspond to a global phase ({\it i.e.}, representing ${\cal R}^{\nu}_{z}[\hat U_0,\hat U_1](\theta_z)$ by a Gaussian SSRC operation) can be expressed as $(n-n_0)\eta=k\pi~\forall ~n$, $k \in \mathbb{Z}^*$. Since $n-n_0 ~\in ~\mathbb{N}^*$, this condition can only be true if $\eta = \pi/p$, with $p \in \mathbb{Z}^*$. However, $\theta_z=(2n_o-N)\eta+2s\pi=(2n_o-N)\pi/p+2s\pi$, $s \in \mathbb{Z}$. This means that if $\hat {\cal R}^{\nu}_{z}[\hat U_0,\hat U_1](\theta_z) \in {\cal G}$, it cannot be a qubit rotation around $z$, {\it i.e.}, ${\cal R}_z(\theta_z)$: given an encoding (that determines the values of $n_0$ and of $N$), all the angles $\theta_z$ cannot be reached within this encoding by applying the SSRC Gaussian operation that was a candidate to  $\hat {\cal R}^{\nu}_{z}[\hat U_0,\hat U_1](\theta_z)$. 

We can alternatively state this result by referring to the physical implementation $\hat H_{\nu}[\hat U_0,\hat U_1]$,  $\hat T_{\nu}[\hat U_0,\hat U_1]$  of the Clifford $\hat H$ (Hadamard gate) and non-Clifford $\hat T$ (phase gate) qubit gates. These gates and their physical implementations, combined, can arbitrarily approach any single qubit operation, including rotations and, in particular, around arbitrary ${\alpha}$ axis. If both $\hat H_{\nu}[\hat U_0,\hat U_1]$ and $\hat T_{\nu}[\hat U_0,\hat U_1]$ could be implemented by SG operations -  rotations -  there would also be a  rotation that would be able to implement any combination of $\hat H_{\nu}[\hat U_0,\hat U_1]$ and $\hat T_{\nu}[\hat U_0,\hat U_1]$, since a combination of rotations is a rotation. In particular, SSRC rotations would be able to arbitrarily approach the effect of any qubit rotation around the $z$ axis (as $\hat H$ and $\hat T$ can arbitrarily reproduce the effect of  ${\cal R}_z(\theta)$). Since this is not the case, we conclude that it is not possible to associate a SG operation to both $\hat H_{\nu}[\hat U_0,\hat U_1]$ and $\hat T_{\nu}[\hat U_0,\hat U_1]$.

Again, for $N=1$ and $n_0=0$ or $n_0=1$ (so $\hat U_0$ and $\hat U_1$ are SG, and we are in the single photon encoding case), the restriction on the values of the phase $\theta_z$ no longer holds, and $\theta_z=\pm \eta + 2s\pi$. 

In conclusion, we have shown that for any encoding, except for the single photon one, the local gates of the universal set must contain at least one SNG gate.


\subsection{The impossibility of building CNOT gates with Gaussian operations}

We now demonstrate that SSRC Gaussian operations (rotations) cannot implement two-qubit entangling gates (such as CNOTs) for any encoding. Consider two qubits encoded in mode partitions $k$ and $k'$, where qubit $k$ is encoded in modes $a_k, b_k$ and qubit $k'$ in modes $a_{k'},b_{k'}$. The logical states of qubit $k$ are encoded as  $\ket{\overline 0}_k=\hat U^{(a_k,b_k)}_0 \ket{N}_{a_k}\ket{0}_{b_k}$, $\ket{\overline 1}_k=\hat U^{(a_k,b_k)}_1 \ket{0}_{a_k}\ket{N}_{b_k}$, where $\hat U^{(a_k,b_k)}_{0(1)}$ represents an arbitrary  (possibly $k$ dependent) unitary operation that may include Gaussian or non-Gaussian transformations involving only the two modes $a_k,b_k$ and that leaves the two possible states of the logical qubits ($\ket{\overline 0}_k$ and $\ket{\overline 1}_k$) orthogonal to one another, ${}_k\langle \overline 0|\overline 1\rangle_k=0$ . The logical states of qubit $k'$ are defined analogously, using operations and modes acting on the partition $k'$. A CNOT gate or equivalent (as a controlled phase gate) implements transformations such as $\ket{\overline 1}_k\ket{\overline 1}_{k'} \to \ket{\overline 1}_k\ket{\overline 0}_{k'}$ in some encoding (as well as the rest of the truth table, but we will focus on this one), which here corresponds to some choice of unitaries $\hat U^{(a_k,b_k)}_{0(1)}$,$\hat U^{(a_{k'},b_{k'})}_{0(1)}$. In we want to implement a CNOT gate with a SSRC Gaussian operation (rotation) we should then have $\hat R \hat U^{(a_k,b_k)}_1\hat U^{(a_{k'},b_{k'})}_1\ket{0}_{a_k}\ket{N}_{b_k} \ket{0}_{a_{k'}}\ket{N}_{b_{k'}} \to \hat U^{(a_k,b_k)}_1\hat U^{(a_{k'},b_{k'})}_0\ket{0}_{a_k}\ket{N}_{b_k} \ket{N}_{a_{k'}}\ket{0}_{b_{k'}}$, where $\hat R$ is a general rotation involving all the four modes $k(i), k'(j)$, $i,j \in \{0,1\}$, that are coupled two by two - a four-mode Gaussian gate. However, 
 \begin{eqnarray}\label{state}
 &&\hat R \hat U^{(a_k,b_k)}_1\hat U^{(a_{k'},b_{k'})}_1\ket{0}_{a_k}\ket{N}_{b_k} \ket{0}_{a_{k'}}\ket{N}_{b_{k'}} = \nonumber \\
 &&(\hat R \hat U^{(a_k,b_k)}_1\hat U^{(a_{k'},b_{k'})}_1\hat R^{\dagger})\hat R \ket{0}_{a_k}\ket{N}_{b_k} \ket{0}_{a_{k'}}\ket{N}_{b_{k'}}=  \nonumber \\
 &&\hat U^{(\tilde a_{ k},\tilde b_{ k})}_1\hat U^{(\tilde a_{ k'},\tilde b_{ k'})}_1\ket{0}_{\tilde a_{ k}}\ket{N}_{\tilde a_{ k}} \ket{0}_{\tilde a_{ k'}}\ket{N}_{\tilde b_{ k'}},
 \end{eqnarray}
 where the symbol tilde stands for a mode basis change, $\hat R \hat U^{(a_k,b_k)}_1\hat R^{\dagger}=\hat U^{(\tilde a_k, \tilde b_k)}_1$, and the same for $k'$,  that does not change the initial state, but rather the mode basis where it is represented. In the general case, for rotations $\hat R$ coupling all the four modes $a_k, b_k$ to modes $a_{k'}, b_{k'}$, the state $ \hat U^{(\tilde a_k, \tilde b_k)}_1 \hat U^{(\tilde a_{k'},\tilde b_{k'})}_1\ket{0}_{\tilde a_k}\ket{N}_{\tilde b_k} \ket{0}_{\tilde a_{k'}}\ket{N}_{\tilde b_{k'}}$ is not separable in the mode partition $(a_k,b_k)$ and $(a_{k'},b_{k'})$. The only scenario where it can be separable in the partition between modes $(a_k, b_k)$ and $(a_{k'},b_{k'})$ is if $\hat R$ is local in each partition $(a_k, b_k)$ and $(a_{k'}, b_{k'})$. In this case, it is separable both in partition $(\tilde a_{k}, \tilde b_k)$ and $(\tilde a_{k'}, \tilde b_{k'})$ and $(a_k, b_k)$ and $(a_{k'}, b_{k'})$. However, if $\hat R$ is local in  modes $(a_k, b_k)$ and $(a_{k'}, b_{k'})$, it cannot implement a two-qubit gate. We conclude that it is impossible to generate a CNOT gate using Gaussian SSRC operations (rotations) alone for any encoding. Non-Gaussian operations are therefore essential for achieving qubit entanglement.



\bibliography{bibCollectiveModes2}

\begin{thebibliography}{139}
\bibitem [{Not({\natexlab{a}})}]{NoteS}%
  \BibitemOpen
  \href@noop {} {\bibinfo {title} {We consider that both modes are equally
  populated because we will take the limit of ${N} \gg 1$.}}
  ({\natexlab{a}})\BibitemShut {NoStop}%
\bibitem [{Not({\natexlab{b}})}]{NoteS2}%
  \BibitemOpen
  \href@noop {} {\bibinfo {title} {One can alternatively displace in phase
  space the squeezed vacuum, inverting the squeezing and the displacement
  operators, but this does not change the general discussion and the fact that
  squeezed states are fully characterized by four parameters.}}
  ({\natexlab{b}})\BibitemShut {NoStop}%
\bibitem [{Com({\natexlab{a}})}]{CommentSqueezing1}%
  \BibitemOpen
  \href@noop {} {\bibinfo {title} {Interestingly, the mode overlap $[\hat c,
  \hat d^{\dagger}]= e^{-2r}$ is directly related to the phase space re-scaling
  induced by squeezing.}} ({\natexlab{a}})\BibitemShut {NoStop}%
\bibitem [{Com({\natexlab{c}})}]{CommentEncoding}%
  \BibitemOpen
  \href@noop {} {\bibinfo {title} {We have used a same ${N}$ for both logical
  states for convenient, but the results can be straightforwardly generalized
  for arbitrary values of the number of photons in each logical qubit.
  {M}oreover, in the {CV} limit, the ${N} \to \infty$ limit is taken, and this
  point becomes irrelevant for states that can considered as balances in terms
  of photon number distribution.}} ({\natexlab{c}})\BibitemShut {NoStop}%
\end{thebibliography}

\end{document}